%% file: main.tex
\documentclass[aps, 
11pt,
letterpaper,
notitlepage,
onecolumn,
nofootinbib,
superscriptaddress,
amsmath,
amssymb,
]{revtex4-1}

\usepackage[english]{babel}
\usepackage[utf8]{inputenc}
\usepackage{amssymb, amsmath, amsthm}
\usepackage{bm,bbm,nicefrac}
\usepackage{color,graphicx}
\usepackage{enumerate}
\usepackage{hyperref}
\usepackage{mathtools}
\usepackage{mathrsfs}
\usepackage{mciteplus}
\usepackage{soul}
\usepackage{xspace}
\usepackage{orcidlink}  
\usepackage{ulem}

\input{shortcuts}

\hypersetup{ 
pdfnewwindow=true,
colorlinks=true,    
linkcolor=cardinal,
citecolor=cardinal,
filecolor=cardinal,
urlcolor=cardinal
} 
 
\begin{document}

%%%%%%%%%%%%%%%%%%%%%%%%%%%%%%%%%%%%
%	Preprint Number
%%%%%%%%%%%%%%%%%%%%%%%%%%%%%%%%%%%%

%%%%%%%%%%%%%%%%%%%%%%%%%%%%%%%%%%%
%   AFFILIATIONS
%%%%%%%%%%%%%%%%%%%%%%%%%%%%%%%%%%%%
\newcommand{\ucb}{Department of Physics, University of California, Berkeley, CA 94720, USA}

\newcommand{\lbnl}{Nuclear Science Division, Lawrence Berkeley National Laboratory, Berkeley, CA 94720, USA}

\newcommand{\boch}{Institut f\"ur Theoretische Physik II, Fakult\"at f\"ur Physik und Astronomie, Ruhr-Universit\"at Bochum, 44780 Bochum, Germany}

\newcommand{\iu}{Physics  Department,  Indiana  University, Bloomington,  IN  47405,  USA}

\newcommand{\ceem}{Center for Exploration of Energy and  Matter, Indiana University, Bloomington, IN 47403,  USA}

%%%%%%%%%%%%%%%%%%%%%%%%%%%%%%%%%%%%
% TITLE
%%%%%%%%%%%%%%%%%%%%%%%%%%%%%%%%%%%%

\title{ On the equivalence of left-hand-cut and three-body formalisms}

%%%%%%%%%%%%%%%%%%%%%%%%%%%%%%%%%%%%
% AUTHORS
%%%%%%%%%%%%%%%%%%%%%%%%%%%%%%%%%%%%

\author{Andr\'e Bai\~ao-Raposo~\orcidlink{0009-0006-5692-2671}}
\email[e-mail: ]{andre.baiaoraposo@ruhr-uni-bochum.de}
\affiliation{\boch}

\author{Raul~Brice\~no~\orcidlink{0000-0003-1109-1473}}
\email[email: ]{rbriceno@berkeley.edu}
\affiliation{\ucb, \lbnl}

\author{Sebastian~M.~Dawid~\orcidlink{0000-0001-8498-5254}}
\email[email: ]{sdawid@iu.edu}
\affiliation{\iu, \ceem}

%%%%%%%%%%%%%%%%%%%%%%%%%%%%%%%%%%%%
%   ABSTRACT
%%%%%%%%%%%%%%%%%%%%%%%%%%%%%%%%%%%%

\begin{abstract}
The ``left-hand cut problem'' in lattice QCD arises when exchange singularities invalidate standard relations used to extract scattering amplitudes from the finite-volume spectrum. In recent years, several approaches were proposed to resolve this problem. We establish analytically and verify numerically the equivalence between the three-body approach proposed in [\hyperlink{https://doi.org/10.1007/JHEP06(2024)051}{JHEP 06 (2024) 051}] and the two-body left-hand cut formalism developed in [\hyperlink{https://doi.org/10.1007/JHEP08(2024)075}{JHEP 08 (2024) 075}]. We test the equivalence in a scalar model describing a particle scattering from an $S$-wave two-particle bound state through one-particle exchange. Both approaches reproduce the same infinite-volume particle--bound-state amplitude and correctly encode the associated left-hand cut in the finite-volume spectrum. Crucially, we identify that the exponentially suppressed finite-volume effects neglected in both formalisms can be numerically large, compromising amplitude extraction at small lattice volumes.
\end{abstract}

\date{\today}
\maketitle

%%%%%%%%%%%%%%%%%%%%%%
% SECTION
%%%%%%%%%%%%%%%%%%%%%%
\section{Introduction}
\label{sec:intro}

Lattice Quantum Chromodynamics (QCD) determinations of two- and three-hadron scattering amplitudes are now well established across a wide range of mesonic and baryonic systems, including resonant and coupled-channel processes~\cite{Dudek:2012xn, Wilson:2014cna, Briceno:2017qmb, Woss:2019hse, Hansen:2020otl, Mai:2021nul, Rodas:2023gma, BaryonScatteringBaSc:2023ori, Wilson:2023anv, Dudek:2024roh, Yan:2024gwp, Erben:2025zph, Dawid:2025doq, Dawid:2025zxc, Briceno:2025yuq, Yan:2025mdm, Lang:2025pjq, Green:2026jgv}. Computing hadronic amplitudes has reached a level of sophistication that requires careful treatment of near-subthreshold singularities, such as left-hand cuts arising from long-range interactions. These structures can significantly affect both finite- and infinite-volume observables~\cite{Green:2021qol, Green:2022rjj, Padmanath:2022cvl, Du:2023hlu, Meng:2023bmz, Collins:2024sfi, Abolnikov:2024key, Prelovsek:2025vbr} and must therefore be systematically incorporated to obtain reliable physical results. Dinucleon systems and the recently discovered $T_{cc}^+(3875)$~\cite{LHCb:2021vvq, LHCb:2021auc} provide important examples of sub-threshold states that lie close to both left- and right-hand cut singularities, and are being actively investigated in lattice QCD~\cite{Padmanath:2022cvl, Chen:2022vpo, Lyu:2023xro, Whyte:2024ihh, Collins:2024sfi, Prelovsek:2025vbr, Alharazin:2026lno, BaryonScattering:2025ziz, BaSc:2026fdy, Detmold:2024iwz, Detmold:2026shh, Zhang:2026cco}.

Several new ways of relating finite-volume energy levels and amplitudes have recently been proposed to address the challenge presented by near-threshold left-hand cuts~\cite{Klos:2016fdb, Meng:2021uhz, Raposo:2023oru, Bubna:2024izx, Hansen:2024ffk, Dawid:2024oey, Raposo:2025dkb, Yu:2025gzg, GomezNicola:2025puj}.
One solution, based on a generic relativistic field theory, was derived in Refs.~\cite{Raposo:2023oru, Raposo:2025dkb}. These works extended the standard derivation of L\"uscher-like finite-volume formalisms~\cite{Luscher:1986pf, Luscher:1991n1, Rummukainen:1995vs, Kim:2005gf, He:2005ey, Briceno:2014oea} by explicitly isolating one-particle exchange diagrams in the diagrammatic expansion of finite-volume four-point correlation functions. They derive a finite-volume quantization condition and infinite-volume integral equations that properly account for the effects of the associated left-hand cut. This formalism connects the finite-volume spectrum to an infinite-volume scheme-dependent function, commonly described as a K matrix. This K matrix is then related to physical scattering amplitudes via a set of integral equations.

Because the formalism is a natural extension of the existing L\"uscher-like methods incorporated in many lattice QCD analyses~\cite{Briceno:2017max, Hansen:2019nir, Davoudi:2020ngi, Mai:2021lwb, Hanlon:2024fjd, Erben:2025zph}, it is a good candidate for numerical implementation in the state-of-the-art computations of hadronic amplitudes. Recent applications~\cite{Dawid:2025wsn, PitangaLachini:2026lyd} illustrate this potential by implementing the formalism to finite-volume spectra in the $T_{cc}^+(3875)$ channel. In particular, Ref.~\cite{PitangaLachini:2026lyd} reported a substantial improvement in the statistical description of the finite-volume spectrum compared to the original analysis~\cite{Whyte:2024ihh}, which used the formalism~\cite{Briceno:2014oea} without the presence of the left-hand cut. Related progress includes an ongoing three-body lattice analysis of the $DD\pi$ system, in which the pion-exchange singularity is treated directly~\cite{Alharazin:2026lno}, and an analysis of the $H$-dibaryon channel using a finite-volume $N/D$ construction that explicitly incorporates the one-pion-exchange left-hand cut~\cite{Rodas:2026zmm}.

This motivates us to perform a closer examination of the formalism. Implementation of this approach requires evaluation of new finite-volume functions and integral equations which differ from the standard L\"uscher method. This complication is familiar from the description of three-particle systems in a finite volume~\cite{Hansen:2014eka, Hansen:2015zga} or from the plane-wave recipe~\cite{Meng:2021uhz} derived from the finite-volume Lippmann-Schwinger equation. However, despite these similarities, each formalism uses conceptually distinct strategies for separating long- and short-range interactions, based on different intermediate kernels and variables. Here we directly compare two approaches by establishing, analytically and numerically, the equivalence of the existing relativistic three-body and two-body left-hand cut formalisms in both infinite and finite volume, in a regime where both apply.

To be concrete, we apply these two formalisms to a physically nontrivial case, in which all particles in the three-body system are identical scalars of mass $m$ and two of them can form an $S$-wave bound state of mass $M$. This model has been previously developed and explored in Refs.~\cite{Romero-Lopez:2019qrt, Jackura:2020bsk, Dawid:2023jrj, Dawid:2023kxu, Briceno:2024txg}, and it exhibits a left-hand cut due to the possible one-particle exchange between a particle and a bound state, similar to one-nucleon exchange in nucleon-deuteron scattering~\cite{Phillips:1969hm, Reiner:1969mmv}. In the kinematic region where one of the particles and the bound state can go on-shell, the resulting two-body scattering amplitude and the finite-volume spectrum should be equally well described by the two formalisms. We prove analytically, up to neglected exponentially suppressed effects, that this is indeed the case.

The equivalence of these two formalisms is not automatic: in the three-body formalism, the bound state is described as an emergent particle appearing as a pole in the two-body subchannel amplitude $\Mc_2$. Information about interactions in this subchannel is not present in the two-body formalism, and it must be translated into additional off-shell contributions to the non-singular kernels that describe particle--bound-state scattering. An important first connection between the infinite-volume components of these formalisms was recently obtained in the nonrelativistic limit~\cite{Dawid:2025wsn}. Here, we achieve that by an explicit reorganization of the three-body equations both in the infinite and finite volume, without relying on the nonrelativistic and ``pole dominance'' approximations of Ref.~\cite{Dawid:2025wsn}.

We then perform numerical investigations of this model. We find strong subpercent numerical evidence of this equivalence for physical energies corresponding to elastic two-body scattering. We analytically continue the numerical results below the particle–bound-state threshold, where the left-hand cut emerges. We find that both formalisms result in amplitudes with consistent left-hand-cut discontinuities, but the agreement is at the few-percent level. The size of the discrepancy is compared to the expected exponentially suppressed finite-volume corrections associated with the particle's mass, the finite size of the two-particle bound state, and proximity to the three-body threshold. We find that these corrections are consistent with the observed difference. We note that, although the L\"uscher formalism was shown to fail in this energy regime~\cite{Dawid:2023jrj} and thus this few percent agreement by far surpasses what can be achieved with the standard quantization condition, this finding may raise a practical concern: exponentially suppressed corrections can become numerically significant and may limit the accuracy attainable with either formalism at currently accessible volumes.

The remainder of this article is organized as follows. Section~\ref{sec:review} reviews the two- and three-body formalisms. Sections~\ref{sec:IV_proof} and \ref{sec:FV_proof} establish their infinite- and finite-volume equivalence, respectively, using the symmetrized OPE throughout. Section~\ref{sec:results} presents the numerical tests, and Sec.~\ref{sec:conclusion} concludes the article with a summary.

%%%%%%%%%%%%%%%%%%%%%%
% SECTION
%%%%%%%%%%%%%%%%%%%%%%
\section{Review of two- and three-body formalisms}
\label{sec:review}

Here, we briefly review the left-hand cut formalism for two-body systems, as presented in Refs.~\cite{Raposo:2023oru, Raposo:2025dkb}  as well as the corresponding three-body formalism presented in Refs.~\cite{Hansen:2014eka, Hansen:2015zga}, both in infinite and finite volume. For the three-particle system, we focus on the amputated scattering amplitude, which is related to the two-body particle–bound-state amplitude via the Lehmann–Symanzik–Zimmermann (LSZ) reduction procedure presented in Ref.~\cite{Jackura:2020bsk}. 

We consider a single-channel elastic two-body scattering of two spinless particles. Although the discussion below can be extended to arbitrary partial waves and spins, for simplicity we assume that only the $S$-wave amplitude contributes, and in most cases we report only the equations satisfied by the $S$-wave amplitudes. The two-body state is composed of a light particle (denoted $\varphi$) with a mass $m$ and a heavy particle with a mass $M$ (a $\varphi \varphi$ bound state denoted $b$). In the numerical investigations presented in Sec.~\ref{sec:results}, we will vary $M$, but it will always satisfy $m < M < 2m$. These particles can interact by exchanging the $\varphi$ particle and through short-range contact interactions. We restrict our attention to the center-of-momentum (c.m.) reference frame. The Mandelstam total invariant mass squared is $s = E^2$, where $E$ denotes the total energy of the system. The relative on-shell momentum of the particles is $\q$, such that $E = \omega(q) + \Omega(q)$ with $\omega(k) = (k^2 + m^2)^{1/2}$ and $\Omega(k) = (k^2 + M^2)^{1/2}$.

%%%%%%%%%%%%%%%%%%%%%%
%%%%%%%%%%%%%%%%%%%%%%
\subsection{Two-body left-hand-cut formalism}
\label{subsec:two-body}

%%%%%%%%%%%%%%%%%%%%%%%%%%%%%%%%%%%%%%%%%%%%%
%%%%%%%%%%%%%%%%%%%%%%%%%%%%%%%%%%%%%%%%%%%%%
\subsubsection{Infinite-volume integral equations}
\label{subsubsec:two-body-IV}

We consider the special case of the formalism in Ref.~\cite{Raposo:2025dkb} adapted to our particular model. The total amplitude for such a system can be written as the sum of two terms\footnote{This is a close analog of the familiar two-potential form of a generic scattering amplitude introduced in Ref.~\cite{Gell-Mann:1953dcn}.}, 
    %%%%%
    \begin{align}
    \label{eq:M_full}
    \Mc_{\rm cut} = \Mc_\Ec+
    \left(1- \Lc\right)
    \left[\Mc_0^{-1} + \Cc\right]^{-1}
    \left(1- \Rc\right).
    \end{align}
    %%%%%
The first of these terms, $\Mc_\Ec$, is the amplitude obtained from summing all so-called ``\textit{ladder diagrams}"; its off-shell version satisfies a Lippmann-Schwinger-like integral equation,
    %%%%%
    \begin{align}
    \label{eq:M_E}
    \Mc_{\Ec,S}(k',k) = \Ec_S(k',k) - \int_p  \, 
    \Ec_S(k',p) \, 
    \Delta_2(p; E) \, 
    \Mc_{\Ec,S}(p,k) \, ,
    \end{align}
    %%%%%
where we used the $S$ subscript to indicate that all quantities have been projected to an $S$ partial wave. We introduced a compact notation for the integral and the measure, which we will use throughout,
    %%%%%
    \begin{align}
    \int_p =  \int_0^{\infty} \frac{dp \, p^2}{2\pi^2} \, .
    \end{align}
    %%%%%
The on-shell $\Mc_{\Ec}$ is obtained in the limit $k,k' \to q$. The one-particle exchange (OPE) contribution is encoded in $\Ec_S$, which is the $S$ partial wave of the $u$-channel propagator,
    %%%%%
    \begin{align}
    \Ec(\k',\k) &= 
    \frac{-g^2}{2}
    \left(
    \frac{1}{\big( \omega(k') - \Omega(k) \big)^2 - (\k' + \k )^2 - m^2  }
    +
    \frac{1}{\big(\omega(k) - \Omega(k') \big)^2 - (\k' + \k )^2 - m^2  } \, 
    \right) \, ,
    \label{eq:E}
    \end{align}
    %%%%%
and can be written as,
    %%%%%
    \begin{align}
    \label{eq:OPE}
    \Ec_S(k',k) &= - \frac{g^2}{2 k' k} \, \frac{1}{2} 
    \Big( Q_0 \big( \zeta(k,k') \big)
    + Q_0\big( \zeta(k',k) \big) \Big) \, , \quad 
    %%%
    \zeta(k,k') = \frac{1}{2k'k} \left[ M^2 - 2 \, \omega(k) \, \Omega(k') \right] \, ,
    \end{align}
    %%%%%
where $Q_0(x)$ is the Legendre function of the 2nd kind and $g$ is the $b \to \varphi \varphi$ coupling in this formalism. Here we enforced $\Ec$ to be symmetric under the interchange of the initial and final states and neglected the $i\epsilon$ prescription, which should be understood implicitly. We note that the on-shell OPE amplitude, $\Ec_S(q,q)$, develops a sub-threshold branch point at
    %%%%%
    \begin{align}
    \label{eq:lhc}
    E_{\rm lhc}^2 = m^2 + 2 M^2 \, .
    \end{align}
    %%%%%
This singularity is inherited by the full amplitude $\Mc_{\Ec,S}$ and starts the OPE left-hand cut.

The two-particle propagator $\Delta_{2}$ includes a pole at the on-shell point, 
    %%%%%
    \begin{align}
    \Delta_{2}(p; E) & = \frac{\omega(q)}{\omega(p)} \, 
    \frac{H_2(p)}{ 2 E \, ( q^2 - p^2 + i \epsilon ) } \, ,
    \label{eq:Delta2}
    \end{align}
    %%%%%
and a momentum-cutoff function $H_2$ which ensures that the integral above converges. To prevent the introduction of power-law finite-volume errors associated with the definition of such a function, one must choose an $H_2$ that is smooth in the region of integration. This criterion excludes step functions, which would amount to a hard cutoff in momentum space. 

In this work, we explored the consequences of two acceptable functions. The first, called ``exp-cutoff'' is of the Gaussian form,
    %%%%%
    \begin{align}
    \label{eq:exp_cutoff}
    H_{2,\rm Exp}(p) = \exp(\alpha \, ( q^2 - p^2 )/m^2 ) \, ,
    \end{align}
    %%%%%
with parameter $\alpha$ controlling the decay width of the smooth regulator. The second considered option, ``error-cutoff,'' is based on the error function,
    %%%%%
    \begin{align}
    \label{eq:error_cutoff}
    H_{2,\rm Err}(p) = 1 + {\rm erf} \left( \alpha (q^2 - p^2)/m^2 \right) \, .
    \end{align}
    %%%%%
In our numerical investigations, we have explored both of these cutoff functions. Although intermediate objects such as $C_S$ or $\Mc_0$ depend on the cutoff, we find the physical conclusions insensitive to the two choices. This is discussed further in Sec.~\ref{sec:results}.

The $S$-wave projection of the short-range part of the amplitude---the second term in Eq.~\eqref{eq:M_full}---is given by
    %%%%%
    \begin{align}
    \Delta \Mc(k',k) = \left( 1 - \Lc_S(k') \right) \left[ \Mc_0^{-1} + \Cc_S  \right]^{-1} \left( 1 - \Rc_S(k) \right) \,,
    \end{align}
    %%%%%
where the endcap functions are
    %%%%%
    \begin{align}
    \label{eq:RcS}
    \Rc_S(k) &= 
    \int_p  \, \Delta_2(p; E) \, \Mc_{\Ec,S}(p,k) \, , \\
    %%%
    \label{eq:LcS}
    \Lc_S(k') &= \int_p  \,
    \Mc_{\Ec,S}(k',p) \, 
    \Delta_2(p; E) \, ,
    \end{align}
    %%%%%
and
    %%%%%
    \begin{align}
    \label{eq:CcS}
    \Cc_S(E) = -\int_p \int_{p'}  \, \Delta_2(p'; E) \, \Mc_{\Ec,S}(p',p) \, \Delta_2(p; E) \, .
    \end{align}
    %%%%%

$\Mc_0$ is a regularization-scheme-dependent ``short-range'' contribution to the total infinite-volume amplitude, i.e. the amplitude governing a hypothetical reaction in which no one-particle exchanges are present. It does not depend on the external momenta $k$ and $k'$ but can depend on the total energy $E$. It has the standard two-body K matrix form,
    %%%%%
    \begin{align}
    \label{eq:M0}
    \Mc_0(E) = \left[ \Kc_0^{-1}(E) - i \rho(E) \right]^{-1} \, ,
    \end{align}
    %%%%%   
where $\rho(E)$ is the two-body phase space for non-identical particles,
    %%%%%
    \begin{align}
    \rho(E) = \frac{q}{8\pi E} \, , \quad q = \frac{1}{2E} \, \lambda^{1/2}(E^2, m^2, M^2) \, ,
    \end{align}
    %%%%%
and $\lambda(x,y,z) = x^2 + y^2 + z^2 - 2(xy+yz+zx)$ is the triangle function. $\Kc_0$ is real and free of the OPE left-hand cut.

Before moving on to the finite-volume component of the formalism, let us make one comment. When solving the integral equation, such as Eq.~\eqref{eq:M_E}, below threshold, one has to take into account moving cuts of the one-particle exchange amplitude entering the integration kernel, $\Ec(k',p)$. These are logarithmic cuts of two Legendre functions in Eq.~\eqref{eq:OPE}, which occur for $\zeta \in [-1,1]$. In the complex momentum space, for fixed $k'$, they are located at
    %%%%%
    \begin{align}
    \label{eq:circ_cut_1}
    p_{{\rm cut},1, \pm}(k', x) &= \frac{M^2}{2 \beta_{x}(k', m) } \, \left[ \, k' x \pm \omega(k') \sqrt{ 1 - \frac{4 \beta_x(k', m)}{M^2} } \, \right] \, , \\
    %%%
    \label{eq:circ_cut_2}
    p_{{\rm cut},2, \pm}(k', x) &= \frac{M^2}{2 \beta_{x}(k',M) } \, \left[ k' x \pm \Omega(k') \sqrt{ 1 - \frac{4m^2 \beta_x(k',M)}{M^4} } \, \right] \, ,
    \end{align}
    %%%%%
where the real parameter $x \in [-1,1]$ and $\beta_x(k', \mu) = k'^2 (1 - x^2) + \mu^2$. Branch points correspond to values $x = \pm 1$. When we set $k' = q$ to obtain the on-shell amplitude, these complex $p$-plane cuts will lead to singularities of $\Mc_{\Ec}$ in the total energy variable, $E$. These happen at energies
    %%%%%
    \begin{align}
    \label{eq:circ_cut_E}
    E_{\rm circ}^{(1)} = \frac{1}{2} \big(M + \sqrt{ 5 M^2 - 4 m^2} \big) \, , \qquad E_{\rm circ}^{(2)} = \frac{1}{m} \big( M^2 - m^2 \big) \, .
    \end{align}
    %%%%%
To solve the integral equation for energies $E < \min(E_{\rm circ}^{(1)}, E_{\rm circ}^{(2)})$, one has to deform the integration contour to avoid cuts~\eqref{eq:circ_cut_1} and~\eqref{eq:circ_cut_2}, as described in Refs.~\cite{Dawid:2023jrj, Dawid:2023kxu}. In this work, we restrict ourselves to higher energies, for which no such 
contour deformation is needed.

%%%%%%%%%%%%%%%%%%%%%%%%%%%%%%%%%%%%%%%%%%%%%
%%%%%%%%%%%%%%%%%%%%%%%%%%%%%%%%%%%%%%%%%%%%%
\subsubsection{Finite-volume quantization condition}
\label{subsubsec:two-body-FV}

 The quantization condition of Refs.~\cite{Raposo:2023oru, Raposo:2025dkb} translates the finite-volume spectrum of a theory enclosed in a periodic cube of side $L$ into a constraint on the infinite-volume $\Mc_0$ amplitude, or equivalently the K matrix defined in Eq.~\eqref{eq:M0}, according to the formula,
    %%%%%
    \begin{align}
    \label{eq:2BQC}
    \det_{\ell m_\ell} \left[ \Mc_0^{-1} + \Cc_L + F \right] = 0 \, .
    \end{align}
    %%%%%
Just like the standard L\"uscher-like approaches, it is a condition for the zeros of a determinant evaluated on matrices acting in the angular momentum basis (hence the $\ell m_\ell$ labels). The quantities appearing in the above condition are:
\begin{enumerate}[(a)]
    \item $\Mc_0$ is the infinite-volume amplitude introduced in Eq.~\eqref{eq:M0},
    \item $\Cc_L$ is an energy-dependent function describing finite-volume power-law contributions to the total amplitude; it is given by the sequence of one-particle exchanges dressed by the external state rescatterings,
    \item $F$ is the difference between the finite- and infinite-volume bubble diagram, describing the power-law effects of the on-shell propagation of the intermediate two-body states.
\end{enumerate}
Due to partial-wave mixing in the box, all objects entering the above formula are matrices in the space of outgoing and incoming two-body angular momentum, $(\ell' m_\ell'; \ell m_\ell)$. Note that $\Mc_0$ is diagonal in this basis, while $\Cc_L$ and $F$ are not. Reference~\cite{Raposo:2025dkb} extended the original formulation of this condition to include particles with nonzero spin and coupled-channel systems. 

The $\Cc_L$ function is a finite-volume version of Eq.~\eqref{eq:CcS} which is formed as a momentum-space product of multiple matrices,
    %%%%%
    \begin{align}
    \label{eq:CL}
    [\Cc_L]_{\ell' m'_\ell ; \ell m_\ell} = 
    - [\Yc^T \, \Delta_{2,L} \,  \Mc_{\Ec,L} \, \Delta_{2,L} \, \Yc ]_{\ell' m'_\ell ; \ell m_\ell} \, .
    \end{align}
    %%%%%
Each object inside the square brackets is a vector or matrix labeled by the finite-volume momenta, $\k' = 2\pi \bm n'/L$, $\k = 2\pi \bm n/L$, with $\bm n', \bm n \in \mathbbm{Z}^3$, such that matrix multiplication denotes summation over the internal momentum index:
    %%%%%
    \begin{align}
    [A B ]_{\k' \k} = \sum_{\p} A_{\k' \p} \, B_{\p \k} \, .
    \end{align}
    %%%%%
We can then define the finite-volume ladder amplitude as
    %%%%%
    \begin{align}
    \Mc_{\Ec,L} = \Ec [\one + \Delta_{2,L} \Ec ]^{-1} \, ,
    \label{eq:MEL}
    \end{align}
    %%%%%
where $[\Ec]_{\k' \k} = \Ec(\k', \k)$ as defined in Eq.~\eqref{eq:E}. The finite-volume two-body Green's function is simply
    %%%%%
    \begin{align}
    [\Delta_{2,L}]_{\k' \k} & = \frac{\delta_{\k' \k}}{L^3} \, \Delta_2(k; E) \, ,
    \label{eq:Delta2L}
    \end{align}
    %%%%%
where $\delta_{\k' \k}$ is the three-dimensional Kronecker delta in discrete momentum variables. These are finite-volume analogs of $\Mc_\Ec$, $\Ec$, and $\Delta_2$, before the partial-wave projection. Finally, the $\Yc$ spherical polynomial vectors allow one to transition between the momentum and angular momentum matrices and are defined as
    %%%%%
    \begin{align}
    [\Yc_{\k}]_{\ell m_\ell} &= \sqrt{4\pi} \, \left( \frac{k}{q} \right)^{\ell} 
    Y_{\ell m_\ell}(\hat \k) \, , \quad 
    %%%
    [\Yc^T_{\k'}]_{\ell' m_\ell'} = \sqrt{4\pi} \, \left( \frac{k'}{q} \right)^{\ell'} 
    Y^*_{\ell' m_\ell'}(\hat \k') \, .
    \label{eq:sph-pol}
    \end{align}
    %%%%%
The last object appearing in the quantization condition is the $F$ function. This is the familiar quantity appearing in standard formulations of the L\"uscher quantization condition, which we define here as 
    %%%%%
    \begin{align}
    [F]_{\ell' m_\ell'; \ell m_\ell} = \left[\frac{1}{L^3} \sum_{\bm k} - \int \frac{d \bm k }{(2\pi)^3} \right]
    \frac{\omega(q)}{\omega(k)} 
    \frac{ [\Yc^T_{\k}]_{\ell' m_\ell'} \, [\Yc_{\k}]_{\ell m_\ell}  }{2 E \, [ q^2 - k^2 + i \epsilon ]} \, .
    \label{eq:FL}
    \end{align}
    %%%%%

In the case of zero angular momenta, $(\ell', m_\ell' ) = (\ell,m_\ell) = (0,0)$, several simplifications occur. Spherical polynomials in the expression for $\Cc_L$ become vectors with all elements equal to unity, which leads to a factorization,
    %%%%%
    \begin{align}
    \label{eq:CL00}
    [\Cc_L]_{00;00} = \Cc_{L,S} = - 
    \sum_{\k', \k}  \, 
    [\Delta_{2,L}]_{\k' \k'} \,
    [\Mc_{\Ec,L}]_{\k' \k} \, 
    [\Delta_{2,L}]_{\k \k} \, .
    \end{align}
    %%%%%
Furthermore, the quantization condition, Eq.~\eqref{eq:2BQC}, becomes an algebraic equation for the $S$-wave $\Kc_{0}$ matrix,
    %%%%%
    \begin{align}
    \Kc_0^{-1} = - \big( \Cc_{L,S} + F_S \big) \, ,
    \label{eq:K0_FV_swave}
    \end{align}
    %%%%%
where $F_S$ is
     %%%%%
     \begin{align}
    F_S = \left[\frac{1}{L^3} \sum_{\bm k} - {\rm PV} \int \frac{d \bm k }{(2\pi)^3} \right]
    \frac{\omega(q)}{\omega(k)} 
    \frac{ 1 }{2 E \, [ q^2 - k^2  ]} \, .
    \label{eq:tildeFL}
    \end{align}
    %%%%%
Given the finite-volume spectrum of the theory, one can recover $\Kc_0$ from Eq.~\eqref{eq:K0_FV_swave}, and, through Eq.~\eqref{eq:M_full}, the full infinite-volume scattering amplitude.

%%%%%%%%%%%%%%%%%%%%%%%%%%%%%%%%%%%%%%%%%%%%%
%%%%%%%%%%%%%%%%%%%%%%%%%%%%%%%%%%%%%%%%%%%%%
\subsection{Three-body formalism}
\label{subsec:three-body}

Now we review the essential aspects of the three-body formalism. We make use of the three-particle integral equations presented in Ref.~\cite{Hansen:2015zga} and the finite-volume quantization condition from Ref.~\cite{Hansen:2014eka}. In the special case containing a two-body bound state, Ref.~\cite{Romero-Lopez:2019qrt} first explained that the three-body formalism should be equivalent to a two-body formalism below the three-body threshold. Ref.~\cite{Jackura:2020bsk} used the LSZ reduction formula to relate the three-body scattering amplitude to the scattering amplitude between the bound state $b$ and the ``spectator" of type $\varphi$, while Ref.~\cite{Dawid:2023jrj} continued the $\varphi b$ amplitude to the left-hand cut region. Here we present integral equations for the $\varphi b$ system in the limit of vanishing three-body K matrix.

%%%%%%%%%%%%%%%%%%%%%%%%%%%%%%%%%%%%%%%%%%%%%
%%%%%%%%%%%%%%%%%%%%%%%%%%%%%%%%%%%%%%%%%%%%%
\subsubsection{Infinite-volume integral equations}
\label{subsubsec:three-body-IV}

We start by reviewing the infinite-volume formalism for the $\varphi b$ scattering amplitude. This amplitude can be written as~\cite{Jackura:2020bsk}
    %%%%%
    \begin{align}
    \label{eq:LSZ}
    \Mc_{\varphi b}(E) = \lim_{k',k \to q} \lambda^2 \, d_S(k',k) \, ,
    \end{align}
    %%%%%
where the $S$-wave amputated \textit{three-body ladder amplitude},
    %%%%%
    \begin{align}
    \label{eq:3B_d_amp}
    d_S(k',k) = -G_S(k',k) - \int_p \frac{1}{2 \omega(p)} \, 
    G_S(k',p) \, 
    \Mc_2(p; E) \, 
    d_S(p,k) \, ,
    \end{align}
    %%%%%
describes a process of subsequent one-particle exchanges between two-body subchannels interacting through the relativistic $\varphi \varphi$ amplitude, $\Mc_2$. In our model, this two-body amplitude is
    %%%%%
    \begin{align}
    \Mc_2(p; E) = \left[ \Kc_2^{-1}(p) - i \rho_2(p) \right]^{-1} \, ,
    \label{eq:M2_amp}
    \end{align}
    %%%%%
in which,
    %%%%%
    \begin{align}
    \Kc_2(p) = - 16 \pi \sqrt{\sigma(p)} \, a \, , \quad  
    \rho_2(p) = \frac{1}{32\pi} \sqrt{1 - \frac{4m^2}{\sigma(p)}} \, ,
    \end{align}
    %%%%%
where the two-body invariant mass squared is $\sigma(p) = (E - \omega(p))^2 - p^2$. For the positive scattering length, $a>0$, the amplitude has a pole at $\sigma_b = M^2 = 4m^2(1 - 1/(ma)^2)$. The residue at the pole is $-\lambda^2$, with $\lambda^2 = 128 \pi M/a$. 

The function $G(\k',\k) = G_{\k' \k}$ describes one-particle exchange (OPE) between $\varphi\varphi$ subsystems, with elements given by
    %%%%%
    \begin{equation}
    G_{\boldsymbol k' \boldsymbol k} = \frac{H_{3} (k') H_{3}(k)}{ \big(E - \omega(k') - \omega(k) \big)^2 - (\boldsymbol k' + \boldsymbol k)^2 - m^2} \,.
    \label{eq:Gmat}
    \end{equation}
    %%%%%
After partial-wave projection to $S$-wave, it takes the form
    %%%%%
    \begin{align}
    \label{eq:three-body-OPE}
    G_S(k',k) = \frac{H_3(k') H_3(k)}{2 k' k} \, Q_0\big(z(k', k) \big) \, ,
    \end{align}
    %%%%%
where 
    %%%%%
    \begin{align}
    z(k',k) = \frac{1}{2 k' k} \left[ \big(E - \omega(k') - \omega(k) \big)^2 - k'^2 - k^2 - m^2 \right] \, .
    \end{align}
    %%%%%
The cutoff function is,
    %%%%
    \begin{align}
    H_3(k) = 
    J \big(\sigma(k)/4m^2 \big) \, , \qquad J(x) = \exp \left( -\frac{1}{x} \exp\left( - \frac{1}{1-x} \right) \right) \, ,
    \end{align}
    %%%%
with $J(x < 0) = 0$ and $J(x > 1) = 1$.\footnote{
Alternatively, one may use a class of ``bumpy'' cutoffs proposed in~\cite{Dawid:2025wsn} which have the property that $H_3(q) = 1$.} The cutoff must be a smooth sigmoid function interpolating between zero and unity to avoid introduction of uncontrolled power-law effects in the finite-volume counterpart of the formalism.

In analogy to the ladder equation in the two-body formalism, to solve the three-body integral equation below the bound-state--particle threshold, one must carefully consider logarithmic cuts of the OPE amplitude, Eq.~\eqref{eq:three-body-OPE}. In particular, for the LSZ-reduced bound-state--particle amplitude, for which we set $k'= k = q$, a circular cut develops in the complex $p$ plane at energies $E < E_{\rm circ}^{(2)}$, see Eq.~\eqref{eq:circ_cut_E}. The presence of this cut requires deformation of the integration contour, as described in Ref.~\cite{Dawid:2023jrj}.

Finally, to compare the two- and three-body formalisms in the next section, it is necessary to remove certain unphysical singularities from $G_S$ in Eq.~\eqref{eq:3B_d_amp}. Namely, following Sec.~3.4 of~\cite{Muller:2021uur}, we decompose
    %%%%%
    \begin{align}
    \label{eq:OPE-decomp}
    G_S(k',k) &= G_{S}^-(k',k) + G_{S}^+(k',k) \, , \\
    %%%
    G_{S}^\pm(k',k) & = \frac{H_{3} (k') H_{3}(k)}{4 k' k} \log\left( \frac{E - \omega(k') - \omega(k) \pm \omega(k'+k)}{E - \omega(k') - \omega(k) \pm \omega(k'-k)} \right) \, .
    \end{align}
    %%%%%
The second term, $G^+_S(k',k)$, contains singularities in the $(k',k)$ plane, appearing only at a sufficiently high momentum scale $k_{\rm sing}$. They do not contribute to the integral equation, since $k_{\rm sing}$ lies outside the support of the cutoff functions. Thus, one can treat $G_{S}^+(k',k)$ as a regular function and absorb it into a redefinition of the short-range three-body interaction. In other words, one may remove $G_{S}^+(k',k)$ from $G_S(k',k)$ and compensate for this by introducing a regularization-scheme-dependent real three-body K matrix in the equation for the three-body amplitude,
    %%%%%
    \begin{align}
    \label{eq:3B_d_amp_modified}
    d_S(k',k) = -\left[ G_S^-(k',k) + \Kc_3(k',k) \right] - \int_p \frac{1}{2 \omega(p)} \, 
    \left[ G_S^-(k',p) + \Kc_3(k',p) \right] \, 
    \Mc_2(p; E) \, 
    d_S(p,k) \, .
    \end{align}
    %%%%%
In what follows, for simplicity, we set $\Kc_3 = 0$. The proof of the equivalence presented in the next section is independent of this assumption, and below we discuss how a non-zero $\Kc_3$ can be included in the derivation.

%%%%%%%%%%%%%%%%%%%%%%%%%%%%%%%%%%%%%%%%%%%%%
%%%%%%%%%%%%%%%%%%%%%%%%%%%%%%%%%%%%%%%%%%%%%
\subsubsection{Finite-volume quantization condition}
\label{subsubsec:three-body-FV}

Below, we again assume that our model is enclosed in a cubic periodic box of side $L$. In the absence of short-range three-particle interactions, and assuming two-particle interactions are $S$-wave dominated, a condition satisfied by the finite-volume spectrum of the three scalar particles is~\cite{Hansen:2014eka}
    %%%%%
    \begin{equation}
    ( F_3^{\rm iso}(E, L) )^{-1} = 0 \,.
    \label{eq:QC3iso}
    \end{equation}
    %%%%%
Here, $F_3^{\rm iso}(E, L)$ is a function defined through
    %%%%%
    \begin{equation}
    F_3^{\rm iso}(E, L) = \sum_{\boldsymbol k', \boldsymbol k} [F_3(E, L)]_ {\boldsymbol k' \boldsymbol k} \, ,
    \label{eq:F3iso}
    \end{equation}
    %%%%%
where $F_3(E, L)$ is a matrix indexed by the incoming and outgoing finite-volume momenta of a chosen spectator, $\boldsymbol k, \boldsymbol k' \in \frac{2\pi}{L} \mathbb{Z}^3$, and can be written explicitly in the form
    %%%%%
    \begin{align}
    F_3 = \frac13 \frac{F_2}{2\omega L^3} - F_2 \frac{1}{1+ \widetilde{\Mc}_{2,L} G} \widetilde{\Mc}_{2,L} F_2 \,,
    \label{eq:F3def}
    \end{align}
    %%%%%
with the different building blocks above also being matrices in this space. $F_2$ is diagonal in the spectator momentum and is closely related to the $F$ function of Eq.~\eqref{eq:FL}, 
    %%%%%
    \begin{align}
    [F_2]_{\k' \k} &= \frac{\delta_{\k' \k}}{2}
    \left[\frac{1}{L^3} \sum_{\bm p} - 
    \int \! \frac{d \bm p }{(2\pi)^3} \right]
    \frac{ H_3(k) }{ 2 \omega(p) \, 2 \omega(|\p - \k|) \, \big(E - \omega(k) - \omega(p) - \omega(|\p - \k|) \big)} \, ,
    \label{eq:F2}
        \\
    &= 
    %%%
    \delta_{\boldsymbol k'\boldsymbol k}
    H_3(k) \, F_{\varphi\varphi}(k;E,L) \, ,
    \label{eq:Fphiphi_def}
    \end{align}
    %%%%%
where we have introduced $F_{\varphi\varphi}$ for future convenience. The matrix $G$ is defined in Eq.~\eqref{eq:Gmat}.

Finally, the elements of $\widetilde{\Mc}_{2,L}$ are given by
    %%%%%
    \begin{align}
    [\widetilde{\Mc}_{2,L}]_{\boldsymbol k' \boldsymbol k}
    & = \frac{ \delta_{\boldsymbol k' \boldsymbol k} }{2\omega(k) L^3} {\Mc}_{2,L}(k; E) \,, \\
    %%%
    {\Mc}_{2,L}(k; E) & = \frac{1}{\Mc_2(k; E)^{-1} + (F_{\varphi \varphi}(k; E, L) + i\rho_2(k) H_3(k) )  } \,,
    \label{eq:M2L}
    \end{align}
    %%%%%
which allows one to recognize ${\Mc}_{2,L}$ as a finite-volume counterpart to the $\varphi\varphi$ amplitude $\Mc_2$.

Using the expression for $\Mc_2$ given in Eq.~\eqref{eq:M2_amp}, one can use the quantization condition of Eq.~\eqref{eq:QC3iso} to obtain the three-body finite-volume spectrum for a given value of the $\varphi\varphi$ scattering length, $a$~\cite{Romero-Lopez:2019qrt}.

%%%%%%%%%%%%%%%%%%%%%%
% SECTION
%%%%%%%%%%%%%%%%%%%%%%
\section{Proof of equivalence of the integral equations}
\label{sec:IV_proof}

Having summarized the key ingredients of these two formalisms in the $J=0$ system of interest, we proceed to prove their equivalence below the three-particle threshold by deriving Eq.~\eqref{eq:M_full} from Eq.~\eqref{eq:LSZ}. The former equation depends on the unspecified regularization-scheme-dependent kernel, $\Kc_0(E)$. Our goal is to show that the three-body formalism can be reduced to a form identical to  Eq.~\eqref{eq:M_full} with a $\Kc_0$ that is a real function of total scattering energy, and does not contain any left-hand singularities. If this is true, then any amplitude that can be described by the three-body framework can be successfully parametrized by the two-body left-hand cut formalism after the appropriate $\Kc_0$ has been chosen. In the following, we pay special attention to regularization factors, without assuming that the regularization conventions used in the two formalisms are the same. 

%%%%%%%%%%%%%%%%%%%%%%%%%%%%%%%%%%%%%%%%%%%%%
%%%%%%%%%%%%%%%%%%%%%%%%%%%%%%%%%%%%%%%%%%%%%
\subsection{Introductory manipulations}

We start by defining an effective $b\to 2\varphi$ coupling~\cite{Dawid:2025wsn},
    %%%%%
    \begin{align}
    g_{\rm eff}^2 = \lambda^2 \, H_3(q)^2 \, .
    \end{align}
    %%%%%
We set $g = g_{\rm eff}$, matching the $b \to \varphi \varphi$ couplings between the two formalisms. For arbitrary off-shell momenta $k'$ and $k$, the symmetrized $S$-wave OPE is given by Eq.~\eqref{eq:OPE}, which we rewrite here for convenience,
    %%%%%
    \begin{align}
    \Ec_S(k',k)
    &=-\frac{g_{\rm eff}^2}{2k'k}\frac{1}{2}
    \left[Q_0\big(\zeta(k,k')\big)+Q_0\big(\zeta(k',k)\big)\right] \, .
    \label{eq:E_sym_proof}
    \end{align}
    %%%%%
To compare this with the three-body exchange, defined in Eq.~\eqref{eq:three-body-OPE} and Eq.~\eqref{eq:OPE-decomp}, define a rescaled three-body OPE amplitude,
    %%%%%
    \begin{align}
    \Gc(k',k)
    &= 
    \frac{- g_{\rm eff}^2 }{H_3(k')H_3(k)} \, G_S^-(k',k) \, .
    \label{eq:C_rescaled}
    \end{align}
    %%%%%
For off-shell momenta, these two are related via the difference,
    %%%%%
    \begin{align}
    \label{eq:epsilon_def}
    \varepsilon(k',k)
    = \Gc(k',k)-\Ec_S(k',k) \, ,
    \end{align}
    %%%%%
By construction, $\varepsilon$ vanishes when both external momenta are on shell, i.e. $\varepsilon(q,q)=0$. Importantly, for off-shell kinematics, $\varepsilon$ is non-singular. Specifically, for $E < 3m$, $\varepsilon(k',k)$, considered here as a function of two real momenta, is analytic in the region $k',k > 0$. This has been ensured by the removal of $G^+_S$ from $G_S$ in Eq.~\eqref{eq:OPE-decomp}.

The two-body subchannel amplitude can be made proportional to the $\Delta_2$ propagator, defined in Eq.~\eqref{eq:Delta2},
    %%%%%
    \begin{align}
    \Mc_2(p;E) &= \left[ -\lambda^2 \, Z(p;E) \right] \, \Delta_2(p;E)\, ,
    \label{eq:M2_Delta2}
    \end{align}
    %%%%%
where
    %%%%%
    \begin{align}
    Z(p;E)
    &=
    \frac{E\omega(p)}{H_2(p)\omega(q)}
    \frac{\sqrt{\sigma(p)}}{M}
    \left[1-\frac{ia}{2}\sqrt{\sigma(p) - 4m^2}\right]
    \left[\frac{q^2-p^2}{\sigma(p) - \sigma(q)}\right] .
    \end{align}
    %%%%%
Given that $\sigma(q) = M^2$, in the on-shell limit the residue function satisfies $Z(q;E) = 2\omega(q)$.

With these new variables, we can rewrite the three-body ladder equation, Eq.~\eqref{eq:3B_d_amp}, as
    %%%%%
    \begin{align}
    \lambda^2 d_S(k',q)
    &=
    \frac{H_3(k') H_3(q)}{H_3(q)^2} \, \Gc(k',q)
    -\int_p
    \frac{H_3(k')H_3(p)}{H_3(q)^2} \, \Gc(k',p) \, 
    \frac{Z(p;E)}{2\omega(p)} \, 
    \Delta_2(p;E) \, 
    \left[\lambda^2 d_S(p,q)\right] .
    \end{align}
    %%%%%
To simplify the notation further, we introduce two more functions $\Wc$ and $\Zc$, defined via 
    %%%%%
    \begin{align}
    \lambda^2 d_S(k',q)
    &=
    \frac{H_3(k') H_3(q)}{H_3(q)^2} \, \Wc(k', q)\, ,
    \\
    \Zc(p) 
    &=
    \frac{Z(p;E)}{2\omega(p)}
    \left(\frac{H_3(p)}{H_3(q)}\right)^2,
    \label{eq:Fcal}
    \end{align}
    %%%%%
where we have left the energy dependence implicit in the definition of $\Zc$. Note that $\Zc(q) = 1$. Another useful observation is that the dependence on $H_3(k')$ can be reabsorbed into the definition of $\Wc$ while not changing the limit in Eq.~\eqref{eq:LSZ}, i.e. $\Wc(k',q) \to \Mc_{\varphi b}(E)$ as $k' \to q$. With these new variables, we can rewrite the ladder equation above as, 
    %%%%%
    \begin{align}
    \Wc(k', q)
    &=
    \Gc(k',q)
    -\int_p \Gc(k',p) \, \Zc(p) \, \Delta_2(p;E) \, \Wc(p, q)\, .
    \end{align}
    %%%%%
At this point, the resemblance between the three-body ladder equation and Eq.~\eqref{eq:M_E}, which defines $\Mc_{\Ec,S}$, becomes more apparent. The important distinctions arise from the difference between the rescaled $\Gc$ and $\Ec_S$ and the fact that $\Zc(p)$ is not equal to $1$ for all momenta. We claim that these differences can be absorbed into the second term appearing in Eq.~\eqref{eq:M_full} with an appropriate choice of the $\Kc_0$ function. To make this evident, let us rewrite the three-body ladder equation one more time by making these differences explicit,
    %%%%%
    \begin{align}
    \Wc(k',q)
    &=
    \big[ \Ec_S(k',q) + \varepsilon(k',q) \big]
    -\int_p
    \big[\Ec_S(k',p)+\varepsilon(k',p)\big]
    \big[\Delta_2(p;E) + \delta_2(p;E)\big] \Wc(p,q)\, ,
    \label{eq:W_def}
    \end{align}
    %%%%%
where we have used the definition of $\varepsilon$ in Eq.~\eqref{eq:epsilon_def} and introduced, 
    %%%%%
    \begin{align}
    \label{eq:delta2}
    \delta_2(p; E) = \Delta_2(p; E) \big[ \Zc(p) - 1 \big] \, ,
    \end{align}
    %%%%%
which is regular at the on-shell point. Physically, $\delta_2$ represents the off-shell difference between the two-body amplitude $\Mc_2$ in the three-body formalism, and the $\Delta_2$ propagator in the two-body left-hand cut approach.\footnote{It is worth noting that the regulating factor in Eq.~\eqref{eq:delta2} makes $\delta_2$ behave as an analog of the Cauchy principal value of $\Delta_2$, in the sense that both $\delta_2$ and ${\rm p.v.} \, \Delta_2$ are real.}

%%%%%%%%%%%%%%%%%%%%%%%%%%%%%%%%%%%%%%%%%%%%%
%%%%%%%%%%%%%%%%%%%%%%%%%%%%%%%%%%%%%%%%%%%%%
\subsection{Recasting the $\varepsilon$ and $\delta_2$ dependence}
\label{sec:operators}
 
We now show how to rewrite the effects of $\varepsilon$ and $\delta_2$ into a new non-singular driving term in the integral equations, which we will call $\Kcz$. To simplify the notation, we will formally treat each building block of the integral equation as an integral operator. Specifically, we define a generic operator $\Ac$ acting in momentum space,
    %%%%%
    \begin{align}
    \Ac = \int_p \int_{p'} \, \Ac(p,p') | p \rangle \langle p' | \, , \qquad \langle p | p' \rangle = \frac{2\pi^2}{p^2} \delta(p - p') \, , \qquad \one = \int_p \, | p \rangle \langle p | \, .
    \end{align}
    %%%%%
The only operators that we will treat as diagonal in momentum space are $\Delta_2$ and $\delta_2$. With this new notation, we can rewrite Eq.~\eqref{eq:W_def} in a more compact form as
    %%%%%
    \begin{align}
    \label{eq:w_int_eq_og}
    \Wc = \Big[ \Ec_S + \varepsilon \Big] - \Big[ \Ec_S + \varepsilon \Big]  \, \Big[ \Delta_2 + \delta_2 \Big] \, \Wc \, .
    \end{align}
    %%%%%
We now show that this expression can be rewritten as
    %%%%%
    \begin{align}
    \Wc = \Big[ \Ec_S + \Kcz \Big] - \Big[ \Ec_S + \Kcz \Big] \, \Delta_2 \, \Wc \, ,
    \label{eq:new_contact}
    \end{align}
    %%%%%
where $\Kcz$, in general, has nontrivial momentum dependence,
    %%%%%
    \begin{align}
    \Kcz = \int_p \int_{p'} \, \Kcz(p,p') | p \rangle \langle p' | \, .
    \end{align}
    %%%%%

To prove this equivalence, we simply match the two expressions and derive the equation that $\Kcz$ must satisfy. Using the symbolic representation above, one can directly solve for $\Wc$ using both identities to find,
    %%%%%%
    \begin{align}
    \Wc &= \Big[ \one + \big[ \Ec_S + \varepsilon \big] \big[ \Delta_2 + \delta_2 \big] \Big]^{-1} \,\big[\Ec_S + \varepsilon \big] \, , 
    \\
    &= \Big[ \one + \big[ \Ec_S + \Kcz \big] \, \Delta_2 \Big]^{-1} \big[\Ec_S + \Kcz \big] \, .   
    \end{align}
    %%%%%%
This equality holds if $\Kcz$ is a solution of the following integral equation,
    %%%%%
    \begin{align}
    \label{eq:wteps}
    \Kcz &= \varepsilon
    - \Big[ \Ec_S + \varepsilon\Big] \, \delta_2 \, \Ec_S -
    \Big[ \Ec_S  + \varepsilon  \Big] \, \delta_2 \, \Kcz \, .
    \end{align}
    %%%%%
Let us note that this equation reproduces the expected answer in the $\delta_2 \to 0$ limit. We also observe that $\Kcz(k',k)$ is a purely real function which can have at most isolated poles in energy.

Indeed, both $\varepsilon(k',k)$ and $\delta_2(k;E)$ are non-singular functions of real momenta. Note that the apparent square-root branch point at $\sigma(k) = 0$, inherited by $\delta_2(k;E)$ from $Z(k;E)$, disappears due to the factor of $H_3(k)^2$ in Eq.~\eqref{eq:Fcal}. This implies that the first term in Eq.~\eqref{eq:wteps} is non-singular. The remaining two terms involve integrals; however, given the properties of $\delta_2(k;E)$, $\varepsilon(k',k)$, and $\Ec_S(k',k)$, they cannot generate any branch points for $E_{\rm circ}^{(2)} < E < 3m$. This is because, for these energies, the integrated functions have no movable (energy-dependent) singularities in the complex momentum plane crossing or pinching the real integration contour, as discussed in detail in Ref.~\cite{Dawid:2023jrj}.

An important result is that the $\Rc_S$, $\Lc_S$, and $\Cc_S$ functions, defined in Eqs.~\eqref{eq:RcS}, ~\eqref{eq:LcS}, and ~\eqref{eq:CcS}, respectively, are all non-singular and therefore real along the left-hand-cut region, $E_{\rm circ}^{(2)} < E < E_{\rm lhc} $. These objects are defined through integrals and have no inhomogeneous terms. One may thus apply to them the same argument that proves that integrals in $\Kc_0^{(t)}$ are non-singular.

With this observation, we note that Eq.~\eqref{eq:new_contact} is essentially the starting point of Refs.~\cite{Raposo:2023oru, Raposo:2025dkb}. In particular, Ref.~\cite{Raposo:2023oru} showed that off-shell momentum-dependent K matrices can be replaced with on-shell momentum-independent K matrices while maintaining sources of left-hand cut singularities isolated solely within the OPE. As a result, the equivalence between these formalisms has been established.

%%%%%%%%%%%%%%%%%%%%%%%%%%%%%%%%%%%%%%%%%%%%%
%%%%%%%%%%%%%%%%%%%%%%%%%%%%%%%%%%%%%%%%%%%%%
\subsection{Final remarks}

To complete our discussion, we compare the two-body formalism with an alternative representation of Eq.~\eqref{eq:new_contact}. To proceed, we rewrite the $S$-wave-projected Eq.~\eqref{eq:M_full} using the operator notation introduced above,
    %%%%%
    \begin{align}
    \label{eq:M_full_v2}
    \Mc_{\rm cut} & = \Mc_{\Ec,S}+
    \left[ \one - \Mc_{\Ec,S} \Delta_2 \right]
    \Tc
    \left[ \one - \Delta_2 \Mc_{\Ec,S} \right] \, , \\
    %%%
    \Tc & =  \Mc_0 + \Mc_0 \, \Delta_2 \Mc_{\Ec,S} \Delta_2 \, \Tc
    \end{align}
    %%%%%
In the two-body formalism, $\Mc_0$ is a rank-1 operator,
    %%%%%
    \begin{align}
    \Mc_0 = \Mc_0(E) \int_p \int_{p'} | p' \rangle \langle p | \, ,
    \end{align}
    %%%%%
where $\Mc_0(E)$ is a momentum-independent function defined in Eq.~\eqref{eq:M0}. Applying standard manipulations to Eq.~\eqref{eq:new_contact}, the corresponding amplitude emerging from the three-body formalism can be brought to a similar form,
    %%%%%
    \begin{align}
    \label{eq:M_full_v3}
    \Wc & = \Mc_{\Ec,S} + 
    \left[ \one - \Mc_{\Ec,S} \Delta_2 \right]
    \Tc^{(t)}
    \left[ \one - \Delta_2 \Mc_{\Ec,S} \right] \, , 
    \\
    %%%
    \mathcal{T}^{(t)} & = \Mcz + \Mcz \, \Delta_2 \Mc_{\Ec,S} \Delta_2 \, \Tc^{(t)} \, .
    \end{align}
    %%%%%
A notable distinction here is that $\Mcz$ has a non-trivial, momentum-dependent kernel,
    %%%%%
    \begin{align}
    \Mcz = \int_p \int_{p'} \Mcz(p',p) \, | p' \rangle \langle p | \, ,
    \end{align}
    %%%%%
and is a solution of the integral equation,
    %%%%%
    \begin{align}
    \Mcz = \Kcz - \Kcz \Delta_2 \Mcz \, ,
    \end{align}
    %%%%%
where $\Kcz$ itself, as given by Eq.~\eqref{eq:wteps}, has a momentum-dependent kernel. 

Let us note that above the $\varphi+b$ threshold, both $\Mc_{\rm cut}$ and $\Wc$ have the right-hand unitarity cut in the $E$ variable, as long as $\Kc_0$ and $\Kcz$ are both real in this region. The cut emerges from integrating the pole of $\Delta_2$, which is identical for both expressions. Furthermore, both amplitudes have the same left-hand cut singularity, given that they share the same inhomogeneous driving term, $\Ec_S \subset \Mc_{\Ec,S}$. The rest of the terms involve integrals, and as argued before, these are purely real in the left-hand cut region. 

The difference between Eqs.~\eqref{eq:M_full_v2} and~\eqref{eq:M_full_v3} arises from the discrepancy between $\Mc_0$ and $\Mcz$, and, indeed, the general off-shell amplitudes $\Mc_{\rm cut}$ and $\Wc$ do not agree for arbitrary external momenta for a given energy. However, a proper choice of real $\Kc_0(E)$ allows one to adjust the on-shell amplitudes to match each other exactly. This can be seen by writing an explicit relation between the two K matrices. Let us define,
    %%%%%
    \begin{align}
    \Delta \Mc^{(t)}(q,q) = \langle q | \big[ \one - \Mc_{\Ec, S} \Delta_2 \big] \,
    \Tc^{(t)} \,
    \big[ \one - \Delta_2 \Mc_{\Ec,S} \big] | q \rangle \, .
    \end{align}  
    %%%%%
Comparing two on-shell amplitudes, Eqs.~\eqref{eq:M_full_v2} and~\eqref{eq:M_full_v3}, we find
    %%%%%
    \begin{align}
    \label{eq:K0-matrix-equivalence}
    \Kc_0^{-1}(E) = \frac{\big(1 - \Lc_S(q) \big) \big(1-\Rc_S(q) \big)}{\Delta \Mc^{(t)}(q,q)} - \Cc_S(E) + i \rho(E) \, .
    \end{align}
    %%%%%
Based on the arguments outlined above, the right-hand side is purely real. Reality for energies $E_{\rm circ}^{(2)} < E < m+M$ is ensured by the fact that each building block is real. Reality along the right-hand cut, $E \geq M + m$, is due to a cancellation of imaginary parts between all terms, which follows from elastic unitarity and can be checked explicitly through a direct calculation; see App.~\ref{app:A}.

Finally, let us address the $\Kc_3 = 0$ assumption from below Eq.~\eqref{eq:3B_d_amp_modified}. A nonzero $\Kc_3$ can be included by replacing $G_S^-$ with $G_S^-+\Kc_3$, assuming $\Kc_3(k',k)$ has a built-in regularization $H_3(k') H_3(k)$. Since $\Kc_3$ is nonsingular in the kinematic region considered, it contributes to $\varepsilon$ as a simple shift, and therefore changes $\Kcz$ in Eq.~\eqref{eq:wteps}, but not the OPE singularities or the steps establishing the equivalence.

%%%%%%%%%%%%%%%%%%%%%%
% SECTION
%%%%%%%%%%%%%%%%%%%%%%
\section{Proof of equivalence of the finite-volume formalisms}
\label{sec:FV_proof}

In this section, we prove the equivalence of the two- and three-body finite-volume formalisms when applied to the model under consideration. We show that, in the kinematic region where both formalisms apply, the finite-volume spectrum predicted by the three-body quantization condition of Eq.~\eqref{eq:QC3iso} is identical to that predicted by the modified two-body condition of Eq.~\eqref{eq:K0_FV_swave}. In the following, we take a broadly similar approach to that used in Ref.~\cite{Briceno:2024txg} to show the equivalence of the three-body quantization condition with the standard L\"uscher quantization applied to the $\varphi b$ system. We assume $\boldsymbol P = 0$ and no short-range three-body interactions, $\Kc_3 = 0$, throughout, as in previous sections, and the total energy below the $3\varphi$ threshold, $E < 3m$.

According to Eq.~\eqref{eq:QC3iso}, the finite-volume energies of the $3\varphi$ system correspond to energies for which the function $F_3^{\rm iso}(E, L)$ diverges. We rewrite this function, previously defined in Eq.~\eqref{eq:F3iso}, in a more convenient form, 
    %%%%%
    \begin{align}
    F_3^{\rm iso} = \sum_{\boldsymbol{k}, \boldsymbol k'}  \left[ \frac13 \frac{F_2}{2\omega L^3} - F_2 \, \widetilde{\Mc}_{2,L} \, F_2 - F_2 \, \widetilde{\Mc}_{2,L} \, d_L \, \widetilde{\Mc}_{2,L} \, F_2 \right]_{\boldsymbol {k}'\boldsymbol k} \,.
    \label{eq:F3iso_v2}
    \end{align}
    %%%%%
Here, we have introduced $d_L$, a finite-volume analog of the ladder amplitude $d_S$, which obeys
    %%%%%
    \begin{align}
    d_L = -G - G \, \widetilde{\Mc}_{2,L} \, d_L \,,
    \label{eq:dL_ladder}
    \end{align}
    %%%%%
i.e.~the analog to Eq.~\eqref{eq:3B_d_amp} before partial-wave projection. On the other hand, the $\varphi b$ spectrum obtained from the two-body formalism satisfies the condition given in Eq.~\eqref{eq:K0_FV_swave}, which we restate here for convenience,
    %%%%%
    \begin{align}
    \Kc_0^{-1} +F_S + \Cc_{L,S} = \Mc_0^{-1} + i\rho + F_S + \Cc_{L,S} = 0 \,.
    \label{eq:2bodyQC_v2}
    \end{align}
    %%%%%
We establish a dictionary between the quantities appearing in each approach.

We start with $F_2$. In our kinematic setting, the invariant mass squared of the $\varphi\varphi$ subsystem is lower than the two-particle threshold, $\sigma(k) = (E - \omega(k))^2 - k^2 < (2m)^2$. Consequently, the L\"uscher function for the $\varphi\varphi$ subsystem, $F_{\varphi\varphi}$, defined in Eq.~\eqref{eq:Fphiphi_def}, cannot have singularities, and all finite-volume corrections must be exponentially suppressed~\cite{Kim:2005gf}. Thus, one can write
    %%%%%
    \begin{align} \nonumber
    [F_2]_{\boldsymbol k', \boldsymbol k} 
    &= 
    \delta_{\boldsymbol k'\boldsymbol k} \, H_{3}(k) \, F_{\varphi\varphi}(k;E,L) \, , \\
    & \simeq 
     -i \delta_{\boldsymbol k'\boldsymbol k} \, H_{3}(k) \, \rho_2(k) \,,
    \label{eq:F2simp}
    \end{align}
    %%%%%
with ``$\simeq$'' meaning ``equal up to terms exponentially suppressed with volume''. The second line follows from the exponential suppression in the volume of the sum-integral difference (with an $i\epsilon$ prescription) below the $\varphi\varphi$ threshold, due to the absence of singularities in the summand/integrand.

Next, we relate $\widetilde\Mc_{2,L}$ and $\Delta_{2,L}$ using definitions introduced in Sec.~\ref{sec:IV_proof}. Neglecting exponentially suppressed differences between $\Mc_{2,L}$ and $\Mc_2$, Eqs.~\eqref{eq:M2L} and \eqref{eq:M2_Delta2} give us
    %%%%%
    \begin{align}
    [\widetilde\Mc_{2,L}]_{\boldsymbol k'\boldsymbol k}
    &\simeq
    -\lambda^2\frac{Z(k;E)}{2\omega(k)}
    [\Delta_{2,L}]_{\boldsymbol k'\boldsymbol k}
    \nonumber\\
    &=-\frac{g_{\rm eff}^2}{H_3(k')H_3(k)}
    \Zc(k)[\Delta_{2,L}]_{\boldsymbol k'\boldsymbol k}\, ,
    \label{eq:M2tildesimp_v1}
    \end{align}
    %%%%%
where we have made use of the finite-volume analog of the $\Delta_2$ pole, defined in Eq.~\eqref{eq:Delta2L}. In the second line, we used $g_{\rm eff}^2=\lambda^2H_3(q)^2$ and the definition of $\Zc$, Eq.~\eqref{eq:Fcal}. Finally, defining the finite-volume analog of $\delta_{2}$, in Eq.~\eqref{eq:delta2},
    %%%%%
    \begin{align}
    [\delta_{2,L}]_{\boldsymbol k'\boldsymbol k}
    &=
    [\Delta_{2,L}]_{\boldsymbol k'\boldsymbol k} \, 
    (\Zc(k)-1), 
    \label{eq:delta2L}
    \end{align}
    %%%%%
we arrive at 
    %%%%%
    \begin{align}
    [\widetilde\Mc_{2,L}]_{\boldsymbol k'\boldsymbol k}
    &\simeq
    -\frac{g_{\rm eff}^2}{H_3(k')H_3(k)}
    [\Delta_{2,L}+\delta_{2,L}]_{\boldsymbol k'\boldsymbol k}\, .
    \label{eq:M2tildesimp}
    \end{align}
    %%%%%

Next, we relate the OPE amplitudes of the two approaches. First, similar to the reasoning in Eq.~\eqref{eq:OPE-decomp}, we remove the unphysical singularities from the OPE amplitude and compensate for this by the introduction of the three-body K matrix, as discussed in Ref.~\cite{Muller:2022oyw}. We split $G$ into two parts, 
    %%%%%
    \begin{align}
    G_{\bm k' \bm k} &= G^+_{\bm k' \bm k} + G^-_{\bm k' \bm k} \, , \\
    %%%%%
    G^\pm_{\bm k' \bm k} &= \mp \frac{1}{2 \omega(|\k+\k'|)} \frac{H_3(k') H_3(k)}{E - \omega(k) - \omega(k') \pm \omega(|\k+\k'|)}
    \end{align}
    %%%%%
As before, the $G^+$ contribution can be absorbed into the short-range three-body interaction, and the remaining physical singularity is preserved in $G^-$. With this, we proceed to treat $G^-$ as the three-body OPE. 

In the spirit of the previous section, we then define the difference between the OPE in the two- and three-body formalisms before partial-wave projection,  
    %%%%%
    \begin{align}
    \varepsilon_{\boldsymbol k'\boldsymbol k}
    &=
    -\frac{g_{\rm eff}^2}{H_3(k')H_3(k)}
    G^-_{\boldsymbol k'\boldsymbol k}
    -\Ec_{\boldsymbol k'\boldsymbol k}\, ,
    \label{eq:epsilon_FV}
    \end{align}
    %%%%%
where $\Ec_{\boldsymbol k'\boldsymbol k}=\Ec(\boldsymbol k',\boldsymbol k)$ is defined in Eq.~\eqref{eq:E}. Equivalently,
    %%%%%
    \begin{align}
    G^-_{\boldsymbol k'\boldsymbol k}
    &=-\frac{H_3(k')H_3(k)}{g_{\rm eff}^2}
    [\Ec+\varepsilon]_{\boldsymbol k'\boldsymbol k}\, .
    \label{eq:Gmatsimp}
    \end{align}
    %%%%%
If one partial-wave projects $\varepsilon_{\boldsymbol k'\boldsymbol k}$, one arrives at the quantity introduced in Eq.~\eqref{eq:epsilon_def}; both quantities vanish when external states go on shell. 

Finally, we consider the finite-volume ladder $d_L$. In the previous section, we introduced an amplitude $\Wc$, proportional to $d_S$, that obeys an integral equation with $\Mc_{\varphi b}$ as its solution. Anticipating a similar relation here, we define a finite-volume amplitude $\Wc_L$ through
    %%%%%
    \begin{equation}
    \lambda^2  [d_L]_{\boldsymbol k'\boldsymbol k} = \frac{H_{3}(k') H_{3}(k)}{H_{3}(q)^2}  [\Wc_L]_{\boldsymbol k'\boldsymbol k} \,,
    \end{equation}
    %%%%%
which reproduces the expected $\Mc_{\varphi b}(E)$ in the on-shell and infinite-volume limits. Using the relations for $\widetilde{\Mc}_{2,L}$ and $G^-$, Eqs.~\eqref{eq:M2tildesimp} and \eqref{eq:Gmatsimp}, the ladder equation of Eq.~\eqref{eq:dL_ladder} becomes
    %%%%%
    \begin{align}
    \Wc_L &= (\Ec + \varepsilon) - (\Ec + \varepsilon) \, [\Delta_{2,L} + \delta_{2,L}] \, \Wc_L \,, 
     \label{eq:FVladder} \\
    &= \sum_{j=0}^\infty (\Ec + \varepsilon) \big[ -  (\Delta_{2,L} + \delta_{2,L}) (\Ec + \varepsilon) \big]^j \,,
    \end{align}
    %%%%%
where the top line can be seen as a finite-volume version of Eq.~\eqref{eq:w_int_eq_og}.

Returning to the expression for $F_3^{\rm iso}$ given in Eq.~\eqref{eq:F3iso_v2}, we will use the results derived above to track and isolate the sources of volume-dependent divergences. Using Eq.~\eqref{eq:F2simp}, we can rewrite the first term of $F_3^{\rm iso}$ as follows:
    %%%%%
    \begin{align}
    \sum_{\boldsymbol{k}, \boldsymbol k'}  \left[ {  \frac{1}{3} } \frac{F_2}{2\omega L^3} \right]
    & \simeq \frac{1}{L^3} \sum_{\boldsymbol k} \left( -\frac{i\rho_2 (k) H_{3}(k)}{6 \omega(k) } \right) \,, \\
    %%%
    & \simeq \int \! \frac{d^3 \boldsymbol k}{(2\pi)^3} \left( -\frac{i\rho_2 (k) H_{3}( k)}{6 \omega(k) } \right)
    \,.
    \label{eq:F3_t1}
    \end{align}
    %%%%%
In the second line, we have replaced the sum with the corresponding integral. We incur only neglected exponentially suppressed corrections in the volume in doing so, since the summand is non-singular. From this result, we immediately see that no finite-volume singularities can arise from this contribution to $F_3^{\rm iso}$.

Therefore, we can write the second and third terms of $F_3^{\rm iso}$ as
    %%%%%
    \begin{align} \nonumber
    & \sum_{\bm{k}, \bm k'} \left[ - F_2 \widetilde{\Mc}_{2,L} F_2 - F_2 \widetilde{\Mc}_{2,L} d_L \widetilde{\Mc}_{2,L} F_2 \right]_{\bm{k}'\bm k} \\ \nonumber
    %%%
    & \simeq -g_{\rm eff}^2 \, \sum_{\bm k, \bm  k'} 
    \rho_2(k')  \, \bigg[ \left(\Delta_{2,L} + \delta_{2,L} \right) - \left(\Delta_{2,L} + \delta_{2,L} \right) \Wc_L \left(\Delta_{2,L} + \delta_{2,L} \right) \bigg]_{\bm k' \bm k} \, \rho_2(k)  
    \\
    %%%
    & = -g_{\rm eff}^2 \, \sum_{\bm k, \bm  k'} 
    \rho_2(k')  \, \bigg[ [\one + (\Delta_{2,L} + \delta_{2,L}) (\Ec + \varepsilon)]^{-1} (\Delta_{2,L} + \delta_{2,L}) \bigg]_{\bm k' \bm k} \, \rho_2(k)   \, , 
    \label{eq:F3_v2}
    \end{align}
    %%%%%
where in the second equality we have used the definition of $\Wc_L$ in Eq.~\eqref{eq:FVladder}. From this representation, it is evident that poles in $\Delta_{2,L}$ do not give rise to poles in $F_3^{\rm iso}$. As a result, the poles of $F_3^{\rm iso}$ can only come from poles $\Wc_L$, as is most evident from the first equality. In other words, the finite-volume spectrum satisfies $\det \Wc_L ^{ -1} = 0$. Applying the two-body finite-volume analysis of Ref.~\cite{Raposo:2025dkb} to ladder equation, Eq.~\eqref{eq:FVladder}, gives Eq.~\eqref{eq:2bodyQC_v2}. This shows that the spectra predicted by the two quantization conditions should be identical up to exponentially suppressed finite-volume corrections. In the numerical study performed in this work, we track three such effects, which we discuss in detail in the next section.

%%%%%%%%%%%%%%%%%%%%%%
% SECTION
%%%%%%%%%%%%%%%%%%%%%%
\section{Numerical investigation }
\label{sec:results}

In this section, we numerically test the equivalence of the two formalisms using the model studied in Refs.~\cite{Romero-Lopez:2019qrt, Jackura:2020bsk, Dawid:2023jrj, Dawid:2023kxu, Briceno:2024txg}. In the three-body system, the only physical parameter\footnote{In principle, the momentum cutoff can be considered as another free parameter, but we keep it fixed.} varied in this study is the two-body scattering length $a>0$. We consider values of $a$ for which a two-body bound state with binding momentum $\kappa = 1/a$ is formed. We assume the isotropic approximation and fix the three-body K matrix to $0$. As previously discussed, we expect these two formalisms to be equivalent up to three classes of exponentially suppressed corrections. 

The first is common to both formalisms, namely, single-particle finite-volume corrections of the order $\mathcal O(e^{-mL})$. Second, the two-body formalism is expected to work well if the volume is large compared to the size of the two-body bound state, i.e., $L\gg 1/\kappa = a$. This is associated with corrections that scale as $\mathcal O(e^{-\kappa L}) = \mathcal O(e^{-L/a})$~\cite{Davoudi:2011md, Briceno:2012rv}. Finally, the two-body formalism breaks down as one approaches the three-particle threshold. These errors are hard to quantify in general, but we draw inspiration from Ref.~\cite{Meissner:2014dea}, and assume them to scale as $\mathcal{O}(e^{-{2\kappa_{\rm th} L}/{\sqrt{3}}})$, where $\kappa_{\rm th} = \sqrt{(3m - E) \, m}$. Note that none of these sources of corrections are directly tied to the presence of the left-hand cut; instead, their relevance is just a consequence of the small volumes for which the finite-volume energies happen to overlap with the left-hand cut in the system considered.

With this in mind, we define three proxies for the expected size of exponentially suppressed effects that we should compare with the residual differences between the finite-volume three-body and left-hand-cut two-body descriptions:
    %%%%%
    \begin{align}
    \sigma_m(E,L) &= e^{-mL} \, ,
    \\
    \sigma_\kappa(E,L) &= e^{-\kappa L} = e^{-L/a} \, ,
    \\
    \sigma_{\rm th}(E,L) &= 
    e^{- 2\kappa_{\rm th} L/\sqrt{3} } \, .
    \end{align}
    %%%%%
We combine them in quadrature to define an approximate diagnostic quantity for the total systematic error,
    %%%%%
    \begin{align}
    \label{eq:sig_sys}
    \sigma_{\rm sys}(E,L) =
    \sqrt{\sigma_m^2(E,L)+\sigma_\kappa^2(E,L)+\sigma_{\rm th}^2(E,L)} \, .
    \end{align}
    %%%%%
Note that in reality, this error is multiplied by unknown prefactors that depend on $m, a, L$ and need not be of order $\Oc(1)$. Generally, for the volume and kinematics considered, $\sigma_\kappa$ is typically the largest of the three in the energy region of the left-hand cut. Moreover, as we discuss below, the observed residual difference between the three-body finite-volume calculation and the left-hand-cut two-body formalism shows dependence on $E$ and $L$ that is qualitatively consistent with $\sigma_{\rm sys}(E, L)$.

%%%%%%%%%%%%%%%%%%%%%%%%%%%%%%%%%%%%%%%%
%%%%%%%%%%%%%%%%%%%%%%%%%%%%%%%%%%%%%%%%
\subsection{Comparing the formalisms}

We compare two formalisms by determining the $\Kc_0(E)$ function via two independent paths:
    %%%%%
    \begin{enumerate}[(1)]
        \item We compute the finite-volume energy levels of the $\varphi b$ system at fixed $a$ and a broad range of volumes using the three-body quantization condition, Eq.~\eqref{eq:QC3iso}. We apply the two-body left-hand cut quantization condition, Eq.~\eqref{eq:K0_FV_swave}, to the extracted spectrum, and solve it for $\Kc_0(E)$ at these energies.
        \item We determine $\Mc_{\varphi b}$ by applying the LSZ reduction to Eq.~\eqref{eq:3B_d_amp}. We match the resulting solution to Eq.~\eqref{eq:M_full}, to determine $\Kc_0$ for any energy through the analog of Eq.~\eqref{eq:K0-matrix-equivalence}, 
            %%%%%
            \begin{align}
            \label{eq:K0_comparison}
            \Kc_0^{-1}(E) = \frac{\big( 1 - \Lc_S(q) \big) \big(1 - \Rc_S(q) \big)}{ \Mc_{\varphi b}(E) - \Mc_\Ec(q,q) } - \Cc_S(E) + i \rho(E) \, .
            \end{align}
            %%%%%
        This requires solving numerically the two-body ladder equation, Eq.~\eqref{eq:M_E}, and evaluating associated integrals in Eqs.~\eqref{eq:RcS},~\eqref{eq:LcS}, and~\eqref{eq:CcS}.
    \end{enumerate}
    %%%%%
The first approach follows steps performed in Refs.~\cite{Romero-Lopez:2019qrt, Briceno:2024txg}, while the second is based on the reasoning of Refs.~\cite{Jackura:2020bsk, Dawid:2023jrj}. In both approaches, we use the same regularization prescription to ensure meaningful comparison of K matrices. Given the discussion in the previous two sections, we expect that $\Kc_0(E)$ obtained in these two ways should be equal up to exponentially suppressed corrections.

%%%%%%%%%%%%%%%%%%%%%%
% FIGURE
%%%%%%%%%%%%%%%%%%%%%%
    \begin{figure}
    \centering
    \includegraphics[width=1.0\linewidth]{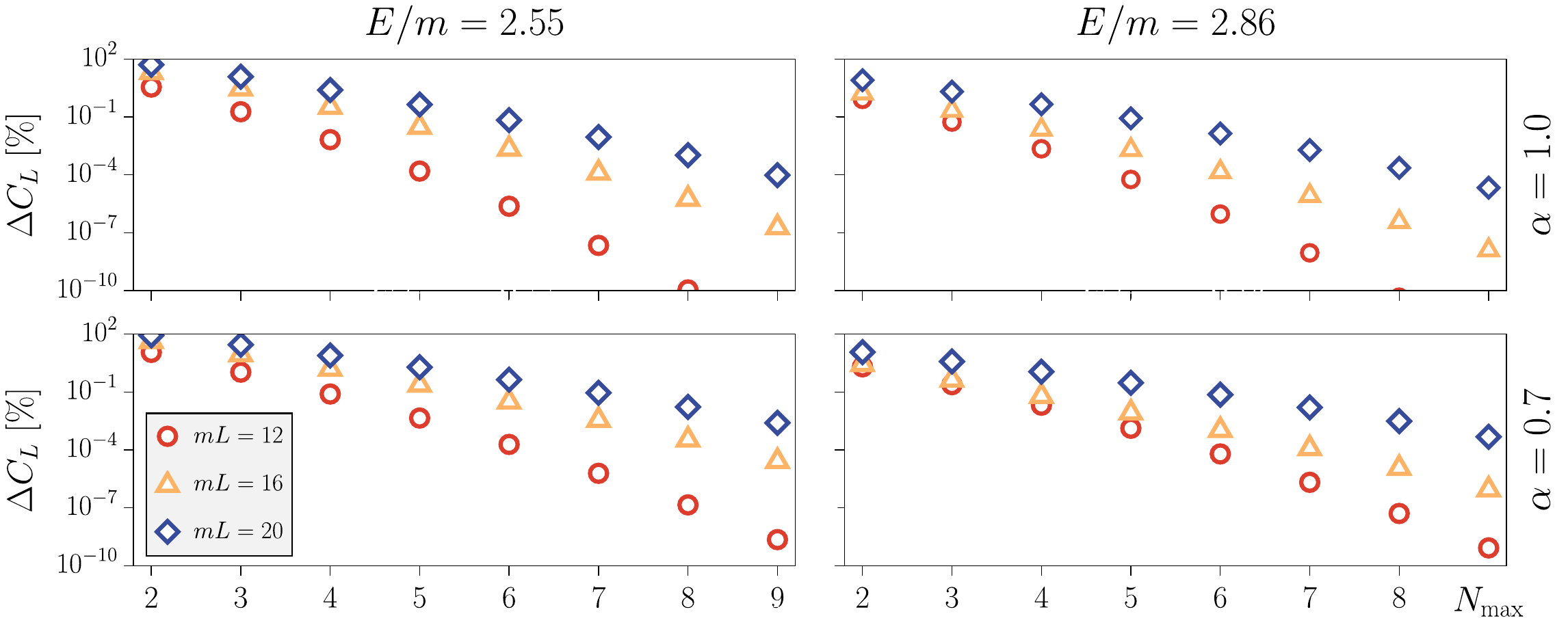}
    \caption{
    The convergence measure, Eq.~\eqref{eq:Delta_CL}, of the finite-volume $C_L$ quantity as a function of the momentum-sum regulator, $N_{\rm max}$, for three different box volumes (see text). We show two energy values: $E/m = 2.55$ (left panels) and $E/m=2.86$ (right panels). We use the exp-cutoff in Eq.~\eqref{eq:exp_cutoff}, with two example values of the regulating parameter: $\alpha=1.0$ (top panels) and $\alpha=0.7$ (bottom panels). 
    }
    \label{fig:CL_converge}
    \end{figure}
%%%%%%%%%%%%%%%%%%%%%%
%%%%%%%%%%%%%%%%%%%%%%

A new key ingredient in the determination of the short-range K matrix from the finite-volume spectrum, as compared to the standard L\"uscher-like methods, is the $S$-wave $\Cc_{L,S}$ function, defined in Eq.~\eqref{eq:CL} for arbitrary partial waves. For a fixed cutoff function, $H_2$, we first solve the algebraic problem in Eq.~\eqref{eq:MEL} by truncating the set of summed momenta at a very high value, $|\k_{\rm max}|$, in physical units. This value is chosen such that $H_2 < 10^{-16}$, which is comparable with the machine precision. Since the matrix sizes grow rapidly with the number of finite-volume momentum shells, in practice, we also define a separate, relatively small, dimensionless integer $N_{\rm max}$ and only include momenta $\k = 2\pi \n/L $, such that,
    %%%%%
    \begin{align}
    \max(|n_x|, |n_y|, |n_z|) \leq N_{\rm max} \, .
    \end{align}
    %%%%%
We insert the solution, $\Mc_{\Ec,L}$, into Eq.~\eqref{eq:CL} and perform the relevant momentum sums defined with the same truncation at $N_{\rm max}$. Finally, we study the convergence rate of $\Cc_{L,S}$ with respect to $N_{\rm max}$. For this purpose, we define
    %%%%%
    \begin{align}
    \label{eq:Delta_CL}
    \Delta C_{L}(N_{\rm max}) = 200 \cdot 
    \frac{ \left| C_{L,S}(N_{\rm max}+1) - C_{L,S}(N_{\rm max}) \right| }{ \left| C_{L,S}(N_{\rm max}+1) + C_{L,S}(N_{\rm max}) \right| } \, . 
    \end{align}
    %%%%%
This quantity is shown in Fig.~\ref{fig:CL_converge}, for several choices of relevant parameters: $E, L$, and $\alpha$. Convergence of $\Cc_L$ as a function of $N_{\rm max}$ is largely independent of the energy, but depends strongly on the smooth cutoff parameter and volume. For larger volumes, a set of finite-volume momenta is finer, and thus a larger $N_{\rm max}$ is needed to include all the relevant momentum shells in the sum. For larger $\alpha$, the number of relevant momentum shells can be suppressed, leading to a mildly faster convergence. We note that the $\alpha$-dependence of the result propagates to the infinite-volume $\Mc_{\Ec}$ and $\Kc_0$ such that the full amplitude, $\Mc_{\rm cut}$, is cutoff independent. We achieve similar conclusions using the error-function cutoff, Eq.~\eqref{eq:error_cutoff}. We find that $N_{\rm max} = 11$ allows one to achieve approximately sub-percent precision for all values of $E$ and $L$.

%%%%%%%%%%%%%%%%%%%%%%
% FIGURE
%%%%%%%%%%%%%%%%%%%%%%
    \begin{figure}
    \centering
    \includegraphics[width=1.0\linewidth]{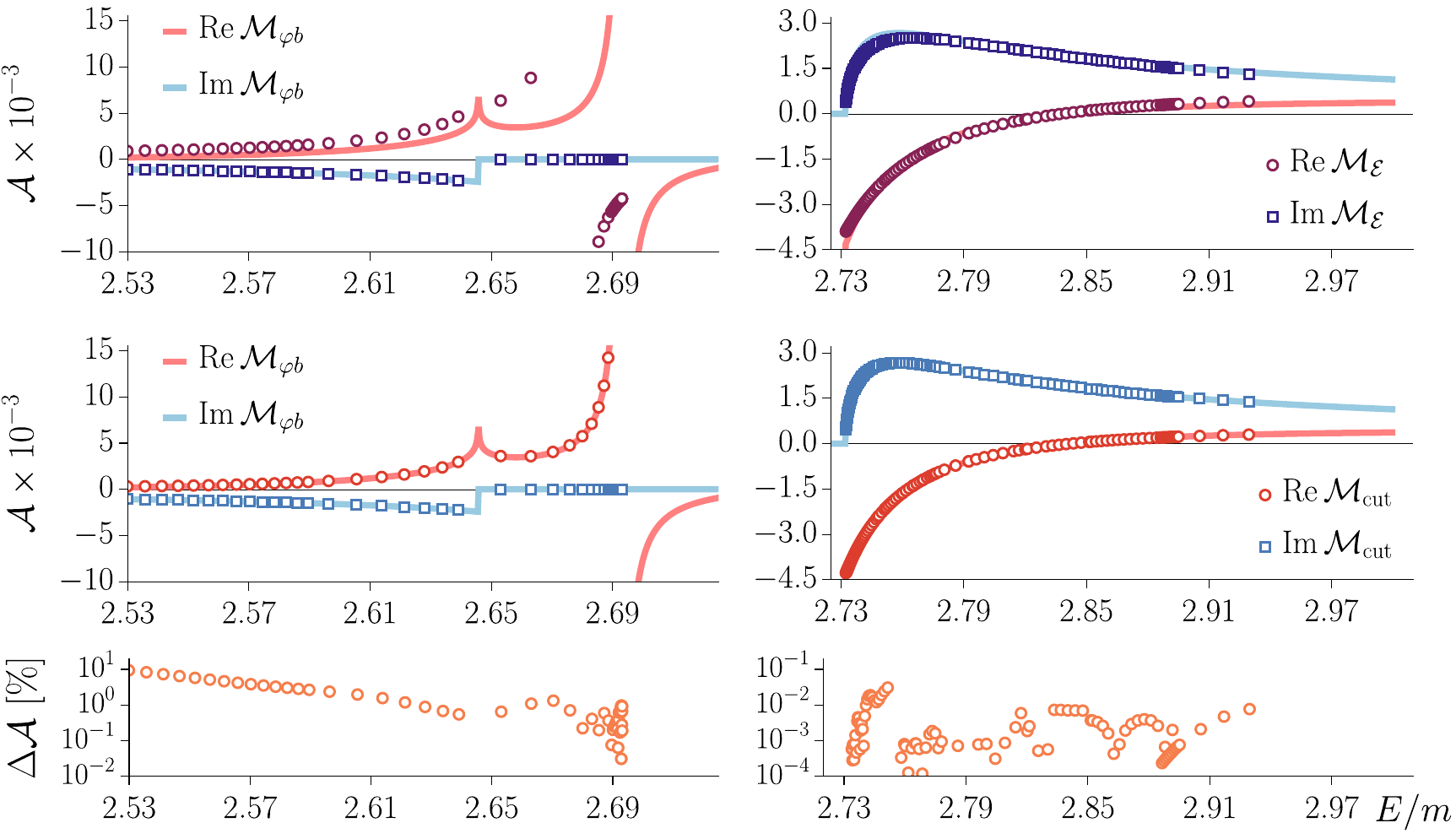}
    \caption{
    Comparison of the particle--bound-state scattering amplitudes, $\Ac$, for $ma=2$. Solid lines are the real and imaginary parts of the solution of the three-body integral equations, $\Mc_{\varphi b}$, reproduced from Ref.~\cite{Dawid:2023jrj}. Circles and squares represent the real and imaginary parts of solutions of the two-body integral equations for the error-cutoff with a regulator $\alpha=1$. The top panel shows a comparison between the three-body solution and the ladder amplitude, $\Mc_{\Ec}$, defined in Eq.~\eqref{eq:M_E}. Note that although the two amplitudes are relatively similar above the threshold, they have a bound-state pole at different positions below the threshold. The middle panel shows the full two-body amplitude, $\Mc_{\rm cut}$, defined in Eq.~\eqref{eq:M_full}. It is obtained from $\Mc_{\Ec}$ by adding the short-range contribution $\Delta \Mc$, with $\Kc_0$ obtained from the quantization condition in Eq.~\eqref{eq:K0_FV_swave}. It is visually indistinguishable from the $\Mc_{\varphi b}$ amplitude and recovers the correct trimer pole. Bottom panels show the residual difference, $\delta \Ac$, defined in Eq.~\eqref{eq:delta_M}.
    }
    \label{fig:amplitudes}
    \end{figure}
%%%%%%%%%%%%%%%%%%%%%%
%%%%%%%%%%%%%%%%%%%%%%

\begin{figure}
    \centering
    \includegraphics[width=1.0\linewidth]{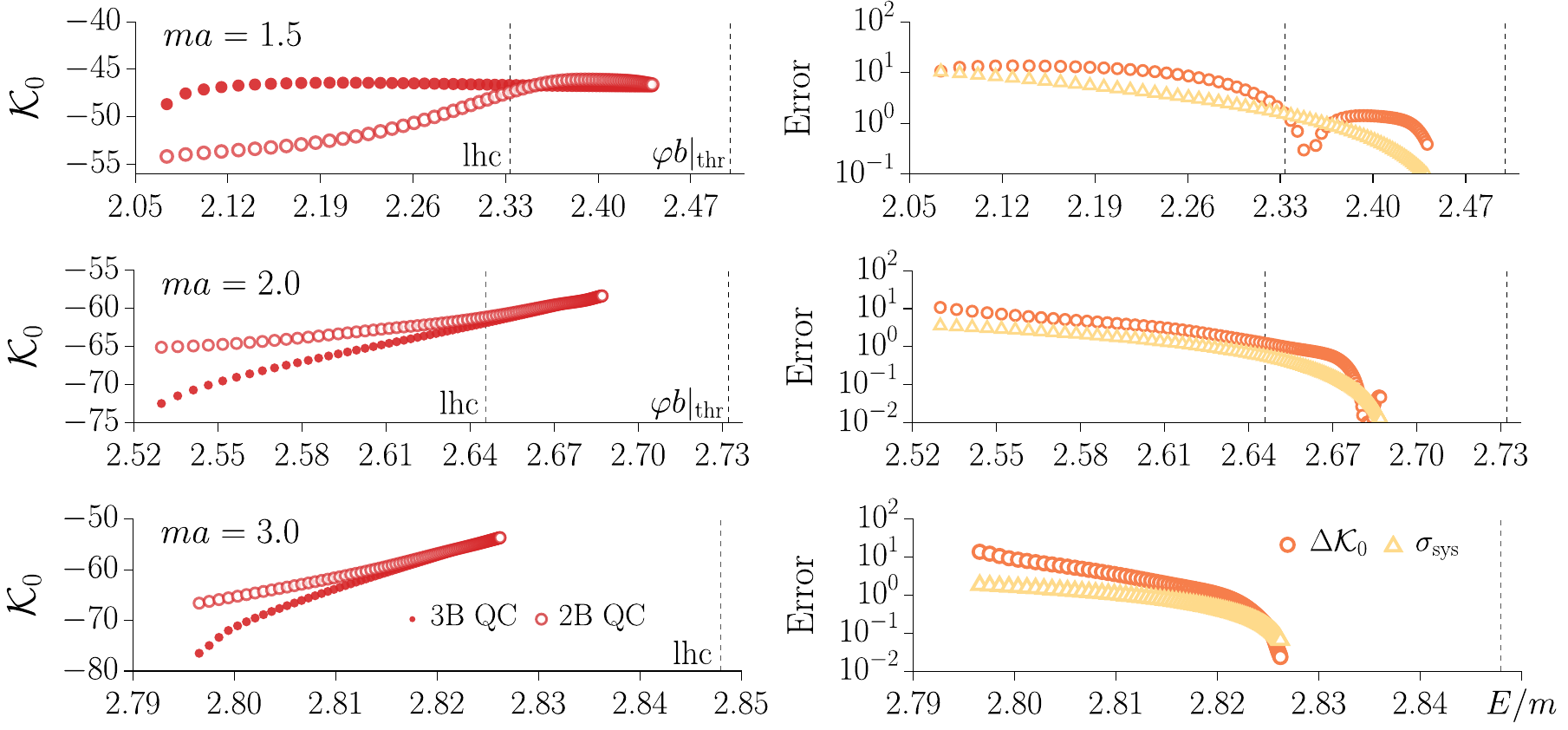}
    \caption{  Comparison of $\Kc_0(E)$ functions obtained from: \textbf{(a)} three-body amplitude through Eq.~\eqref{eq:K0_comparison} (``3B QC'', filled points) and \textbf{(b)} two-body quantization condition, Eq.~\eqref{eq:2BQC} (``2B QC'', open points). We used the exp-cutoff with $\alpha=1$. The right panels show results for three different scattering lengths, $ma = 1.5, 2.0, 3.0$. The left panels show the related relative difference between the two results, and compare them to the estimate of the leading exponential finite-volume effects.}
    \label{fig:K0-comparison}
\end{figure}

We present the numerical comparison of the formalisms in two ways. In Fig.~\ref{fig:amplitudes}, we show the differences between the scheme-independent physical amplitudes obtained using methods (1) and (2) at a fixed value of the scattering length, $ma = 2$. For this computation, we chose the error-cutoff with parameter $\alpha = 1$. In Fig.~\ref{fig:K0-comparison}, we present the differences between the scheme-dependent K matrices obtained using three different values of the scattering length, $ma = 1.5, 2.0,$ and $3.0$. For this computation, we used exp-cutoff. Corresponding plots obtained with other types of regularizations exhibit similar behavior and are not shown. To compare the quantities obtained with methods (1) and (2), we introduce a relative error between amplitudes,
    %%%%%
    \begin{align}
    \label{eq:delta_M}
    \Delta \Ac(E) = 200 \cdot 
    \frac{ \left| \Mc_{\varphi b}(E) - \Mc_{\rm cut}(E) \right| }{ \left| \Mc_{\varphi b}(E) + \Mc_{\rm cut}(E)  \right| } \, . 
    \end{align}
    %%%%%
For the K matrices, the error, $\Delta \Kc_0$, is obtained through the analogous formula, with amplitudes replaced by K matrices obtained using the two- and three-body formalisms.

There are two notable features of Fig.~\ref{fig:amplitudes} that are worth highlighting. First, we observe that although solutions of the two- and three-body ladder equations produce similar amplitudes, they do not agree exactly, especially in the region of the three-body bound-state pole. Indeed, without a non-zero $\Kc_0$, both models have different spectra. It is not surprising, since the $\Mc_{\Ec}$ amplitude carries a non-trivial regularization dependence, which can only be compensated by an appropriate $\Delta \Mc$ term. When this term is included, as shown in the middle panel, the amplitudes match nearly perfectly, and the three-body pole position is recovered. Moreover, the full amplitude shows a much better agreement with the three-body result in the left-hand cut region. This may be compared with Fig.~12 of~Ref.~\cite{Dawid:2023jrj}, where a significantly larger deviation occurs between the solution of the three-body calculation and the standard L\"uscher formalism. This shows that the two-body left-hand cut quantization condition reproduces correct physics and that the proposal of Refs.~\cite{Raposo:2023oru, Raposo:2025dkb} constitutes a clear improvement of the standard two-body framework in the generic model under study.

%%%%%%%%%%%%%%%%%%%%%%
%%%%%%%%%%%%%%%%%%%%%%
%%%%%%%%%%%%%%%%%%%%%%
%%%%%%%%%%%%%%%%%%%%%%

Second, we note that above the threshold, the amplitudes match with sub-percent-level precision. However, the agreement between the two results worsens in the deeper sub-threshold energy region, reaching a tension of order 10\%. We associate this with the growing importance of the exponentially suppressed effects that are ignored in the derivation of both the two- and three-body formalism. This can be seen in Fig.~\ref{fig:K0-comparison}, where we present a result of computing K matrices using methods (1) and (2). We see that the discrepancy between these objects grows with decreasing energy, which follows from the fact that lower-energy data in our model occur at smaller $L$, where these exponential effects become larger. For instance, at $ma=2$, finite-volume energies from below the OPE branch point are obtained for box sizes $mL \in [6.5, 11]$, with the lower value corresponding to lower energy. The corresponding ratio $L/a \in [3.25, 5.50]$. One might na\"ively expect that decreasing $ma$ should improve the ratio $L/a$ and thus the systematic error; however, interestingly, for this particular system, the subthreshold energy levels overlapping with the left-hand cut are obtained for low volumes at low scattering lengths and for large volumes as $ma$ increases. For $ma=1.5$, energies are obtained for $mL \in [2, 6]$, which leads to $L/a \in [1.3, 4.0]$. For $ma=3.0$, energies are obtained for $mL \in [12, 40]$, which yields $L/a \in [4.0, 13.3]$. We see that the error rapidly decreases as the volume increases, visually agreeing with the behavior of the diagnostic $\sigma_{\rm sys}$ we introduced in Eq.~\eqref{eq:sig_sys}.

These results demonstrate that the exponentially suppressed finite-volume effects—neglected in both formalisms—can nevertheless be numerically large, especially in the presence of shallow bound states, as in the model we study. This potentially limits amplitude extractions at small lattice volumes using the two-body formalism. It is important to emphasize that for such shallow bound-state scenarios, the three-body formalism would not suffer from errors that scale as $e^{-\kappa L}$. We stress also that we do not see evidence that these corrections are due to the left-hand cut itself, but instead due to the small box volumes for which the finite-volume energies happen to be on the cut.

%%%%%%%%%%%%%%%%%%%%%%
% SECTION
%%%%%%%%%%%%%%%%%%%%%%
\section{Conclusion}
\label{sec:conclusion}

We have established the equivalence between the relativistic three-body formalism~\cite{Hansen:2014eka, Hansen:2015zga} and the two-body left-hand-cut formalism~\cite{Raposo:2023oru, Raposo:2025dkb} in a kinematic region where the two-particle subsystem forms a bound state and both descriptions apply. By reorganizing the three-body integral equations and their finite-volume counterparts, we showed that the particle--bound-state amplitude and spectrum can be expressed in the two-body form. This provides a nontrivial connection given differences between the original formalisms. Both approaches are capable of describing systems driven by long-range one-particle exchanges; in doing so, the short-distance function $\Kc_0$ of the two-body formalism absorbs the nonsingular off-shell contributions and reproduces correct physics. We discussed how this K matrix can be reproduced from the three-body approach.

We supplemented this analytic proof with a numerical investigation that provides a test of these relations. We used the three-body quantization condition to obtain the finite-volume spectrum below the three-body threshold. Then we applied the finite-volume two-body formalism to these energy levels to predict the $\Kc_0$ matrix. Independently, we solved for $\Kc_0$ by matching the infinite-volume amplitudes of these two formalisms.

Above the particle-bound-state threshold, the two determinations of $\Kc_0$ agree at the subpercent level. After continuation onto the left-hand cut, agreement is at the few-percent level. The remaining differences are compatible with our estimate of the exponentially suppressed finite-volume effects omitted in establishing the equivalence, governed by the particle's mass, the bound-state scale, and the proximity to the three-particle threshold. The observed tension grows precisely in the region where these exponentially suppressed terms are no longer small due to the shrinking box size. We note that related issues have been investigated previously in the context of $NN$ scattering~\cite{Bedaque:2006yi, Sato:2007ms}, but further systematic studies would be valuable.

These finite-volume approaches are being applied in novel studies of interesting physical systems, such as $T_{cc}^+(3875)$~\cite{PitangaLachini:2026lyd, Alharazin:2026lno} and dinucleon scattering~\cite{BaSc:2026fdy}. Together with Ref.~\cite{Dawid:2025wsn}, our analysis provides explicit relations between K matrices and quantization conditions, and could help clarify systematic differences arising from differences in both descriptions, such as differing regularization schemes and parametrizations of K matrices. Assessing how these and other finite-volume effects propagate to extraction of hadronic amplitudes may prove useful in application to lattice QCD spectra.

%%%%%%%%%%%%%%%%%%%%%%
\section*{Acknowledgments}
%%%%%%%%%%%%%%%%%%%%%%

We thank Maxwell Hansen, Andrew Jackura, Samantha Goldberg, Nelson Pitanga Lachini, Fernando Romero-L\'opez, and Stephen Sharpe for useful discussions. We also thank Fernando Romero-L\'opez for sharing finite-volume energy levels from Ref.~\cite{Romero-Lopez:2019qrt}. ABR is supported by the European Research Council (ERC) consolidator grant StrangeScatt-101088506. RAB was partly supported by the U.S. Department of Energy, Office of Science, Office of Nuclear Physics under Award No. DE-SC0025665 and No. DE-AC02-05CH11231. This work contributes to the goals of the USDOE ExoHad Topical Collaboration, contract DE-SC0023598.

\appendix

%%%%%%%%%%%%%%%%%%%%%%
\section{Reality of the short-range K matrix}
\label{app:A}

In this section we prove that the imaginary part of Eq.~\eqref{eq:K0-matrix-equivalence} vanishes for energies above the $\varphi b$ threshold, $E \geq M + m$, such that $\Kc_0$ does not contain a right-hand cut. For convenience, we reproduce this equation here,
    %%%%%
    \begin{align}
    \label{eq:K0-matrix-equivalence-2}
    \Kc_0^{-1}(E) = \frac{\big(1 - \Lc_S(q) \big) \big(1-\Rc_S(q) \big)}{\Delta \Mc^{(t)}(q,q)} - \Cc_S(E) + i \rho(E) \, .
    \end{align}
    %%%%%
The derivation proceeds through repeated application of the product formula 
    %%%%%
    \begin{align}
    \im[AB] = \im[A] B + A^* \im{B} \, ,
    \end{align}
    %%%%%
and the off-shell unitarity relation for $\Mc_{\Ec}$,
    %%%%%
    \begin{align}
    \im \Mc_{\Ec}(k',k) = \Mc_{\Ec}^*(k',q) \rho(q) \Mc_{\Ec}(q,k) \, ,
    \end{align}
    %%%%%
Unitarity also implies that $\xi = 1 + 2 i \rho(q) \Mc_{\Ec}(q,q)=e^{2i\delta_{\Ec}}$ where $\delta_{\Ec}$ is the phase shift of $\Mc_{\Ec}$. Therefore, $|\xi|=1$. The same is true for $\eta = 1 + 2 i \rho(q) \Wc(q,q) = e^{2i\delta_{\Wc}}$. 

Since the OPE amplitude is symmetric, so is $\Mc_{\Ec}(k',k)$, which implies $\Lc_S(q) = \Rc_S(q)$. Moreover, we can show that
    %%%%%
    \begin{align}
    \label{eq:appA1}
    \big(1 - \Lc_S(q) \big) \big(1-\Rc_S(q) \big) = \xi \, |1-\Rc_S(q) |^2 \, .
    \end{align}
    %%%%%
This follows from $z = z^* + 2 i \im z$ and the calculation:    
    %%%%%
    \begin{align}
    \im (1 - \Lc_S(q)) &= \im \left[ 1 - \int_p \Delta_2(p) \Mc_{\Ec}(p,q) \right] \\
    & = - \int_p \im \big[ \Delta_2(p) \big] \Mc_{\Ec}(p,q) - \int_p \Delta_2^*(p) \im \Mc_{\Ec}(p,q) \\
    & = \rho(q) \Mc_{\Ec}(q,q) \left[ 1 - \int_p \Delta_2^*(p) \Mc_{\Ec}^*(p,q) \right] \, .
    \end{align}
    %%%%%
Thus,
    %%%%%
    \begin{align}
    1-\Rc_S(q)  = \big(1 + 2 i \rho(q) \Mc_{\Ec}(q,q) \big) \, \big( 1-\Rc_S^*(q) \big) \, ,
    \end{align}
    %%%%%
which proves Eq.~\eqref{eq:appA1}, given the equality between $\Lc_S$ and $\Rc_S$.

Furthermore,
    %%%%%
    \begin{align}
    \label{eq:appA2}
    \im \big[ \Cc_S - i \rho \big] = - \rho(q) \, |1-\Rc_S(q) |^2 \, .
    \end{align}
    %%%%%
Indeed, using the definition of Eq.~\eqref{eq:CcS},
    %%%%%
    \begin{align} \nonumber
    \im \Cc_S & = - \int_p \int_{p'} \im[ \Delta_2(p')] \Mc_{\Ec}(p',p) \Delta_2(p) 
    - \int_p \Delta_2^*(p') \im[\Mc_{\Ec}(p',p)] \Delta_2(p) \\ 
    %%%
    & \quad \, - \int_p \Delta_2^*(p') \Mc_{\Ec}^*(p',p) \im[\Delta_2(p)] \\ \nonumber
    %%%
    & = \rho(q) \int_p \Mc_{\Ec}(q,p) \Delta_2(p) 
    - \int_p \Delta_2^*(p') \Mc_{\Ec}^*(p',q) \rho(q) \Mc_{\Ec}(q,p) \Delta_2(p) \\ 
    %%%
    & \quad \, + \rho(q) \int_{p'}\Delta_2^*(p') \Mc_{\Ec}^*(p',q) = \rho(q) \big[ \Lc_S(q) - \Lc_S(q) \Rc_S^*(q) + \Rc_S^*(q) \big] \, .
    \end{align}
    %%%%%
Thus, again, given that $\Lc_S(q) = \Rc_S(q)$,
    %%%%%
    \begin{align}
    \im \big[ \Cc_S - i \rho \big] = - \rho(q) \big[ 1 - \Lc_S(q) - \Rc_S^*(q) + \Lc_S(q) \Rc_S^*(q) \big] = - \rho(q) \, |1-\Rc_S(q) |^2 \, .
    \end{align}
    %%%%%

Results~\eqref{eq:appA1} and~\eqref{eq:appA2} can now be used in the final derivation. We rewrite,
    %%%%%
    \begin{align}
    \Kc_0^{-1}(E) = 2 i \rho(q) \frac{\xi}{\eta - \xi} |1-\Rc_S(q) |^2 
    - \Cc_S(E) + i \rho(E)
    \end{align}
    %%%%%
Here, we replaced $\Delta \Mc^{(t)}(q,q) = \Wc(q,q) - \Mc_{\Ec}(q,q)$ in the denominator of the first term, and multiplied numerator and denominator by $2 i \rho(q)$. The imaginary part of $\Kc_0^{-1}$ is then
    %%%%
    \begin{align}
    \im \Kc_0^{-1}(E) & \propto \im [2 i \rho(q)] \, \frac{\xi}{\eta - \xi} - 2 i \rho(q) \im\left[ \frac{\xi}{\eta - \xi} \right] + \rho(q) \\
    %%%
    & = \rho(q) \, \left[ \frac{\eta}{\eta - \xi} + \frac{\xi^*}{\eta^* - \xi^*} \right] = 0 \, .
    \end{align}
    %%%%%
In the final line, we used $|\eta| = |\xi| = 1$.

%%%%%%%%%%%%%%%%%%%%%%
% BIBLIOGRAPHY
%%%%%%%%%%%%%%%%%%%%%%
\bibliographystyle{apsrev4-1}
\bibliography{main}

\end{document}

%% file: shortcuts.tex
\renewcommand{\k}{\bm{k}}
\newcommand{\p}{\bm{p}}
\newcommand{\q}{\bm{q}}

\newcommand{\n}{\bm{n}}

\newcommand{\one}{\mathbbm{1}}

\newcommand{\Ac}{\mathcal{A}}

\newcommand{\Cc}{\mathcal{C}}

\newcommand{\Ec}{\mathcal{E}}

\newcommand{\Gc}{\mathcal{G}}

\newcommand{\Kc}{\mathcal{K}}
\newcommand{\Kcz}{\mathcal{K}_0^{(t)}}
\newcommand{\Mcz}{\mathcal{M}_0^{(t)}}

\newcommand{\Lc}{\mathcal{L}}
\newcommand{\Mc}{\mathcal{M}}

\newcommand{\Oc}{\mathcal{O}}

\newcommand{\Rc}{\mathcal{R}}

\newcommand{\Tc}{\mathcal{T}}

\newcommand{\Wc}{\mathcal{W}}

\newcommand{\Yc}{\mathcal{Y}}
\newcommand{\Zc}{\mathcal{Z}}

\DeclareMathOperator{\im}{Im}

\renewcommand{\min}{{\text{min}}}

\newcommand{\bcol}{\left[ \begin{array}{c}}
\newcommand{\ecol}{\end{array} \right]}
\newcommand{\beq}{\begin{eqnarray}}
\newcommand{\eeq}{\end{eqnarray}}

\newcommand{\kev}{\ensuremath{{\mathrm{\,ke\kern -0.1em V}}}\xspace}
\newcommand{\mev}{\ensuremath{{\mathrm{\,Me\kern -0.1em V}}}\xspace}
\newcommand{\gev}{\ensuremath{{\mathrm{\,Ge\kern -0.1em V}}}\xspace}
\newcommand{\tev}{\ensuremath{{\mathrm{\,Te\kern -0.1em V}}}\xspace}

\usepackage{mfirstuc} % to uppercase the name
\newcommand{\addReviewer}[2]{
  \expandafter\newcommand\csname #1\endcsname[1]{{\sf \color{#2} {#1}:\,##1}}
  \expandafter\newcommand\csname #1cor\endcsname[2]{{\color{#2} {#1}:\,\st{##1}{\sf ##2}}}
  \expandafter\newcommand\csname #1color\endcsname{#2}
}

\usepackage{soul,color}
\definecolor{chromeyellow}{rgb}{1.0, 0.65, 0.0}
\definecolor{DodgeBlue}{rgb}{0.118, 0.565,1.000}
\definecolor{asparagus}{rgb}{0.53, 0.66, 0.42}
\definecolor{cadmiumgreen}{rgb}{0.0, 0.42, 0.24}
\definecolor{cardinal}{rgb}{0.77, 0.12, 0.23}
\definecolor{jlab_red}{RGB}{192,39,45}
\definecolor{jlab_orange}{RGB}{249,102,0}
\definecolor{jlab_blue}{RGB}{47,122,121}
\definecolor{jlab_green}{RGB}{65,125,10}
\definecolor{bobcat_green}{RGB}{2,66,48}

\addReviewer{ab}{jlab_green}
\addReviewer{rb}{jlab_blue}
\addReviewer{sebastian}{cardinal}